\documentclass[aps,prb,twocolumn,showpacs,amsmath]{revtex4}
\usepackage{latexsym}
\usepackage{amssymb}
\usepackage{graphicx}
\usepackage{bm}
\usepackage[english]{babel}
\usepackage[latin1]{inputenc}
\usepackage{amsmath}
\usepackage{amsfonts}

\newcommand{\pdag}{{\phantom{\dagger}}}
\newcommand{\bq}{\begin{equation}}
\newcommand{\eq}{\end{equation}}
\newcommand{\bn}{\begin{eqnarray}}
\newcommand{\en}{\end{eqnarray}}

\begin{document}

\title{Full counting statistics of phonon-assisted Andreev tunneling through a quantum dot coupled to normal and superconducting 
leads}

\author{Bing Dong}
\affiliation{Key Laboratory of Artificial Structures and Quantum Control (Ministry of Education), Department of Physics and 
Astronomy, Shanghai Jiaotong University, 800 Dongchuan Road, Shanghai 200240, China}
\affiliation{Collaborative Innovation Center of Advanced Microstructures, Nanjing, China}

\author{G. H. Ding}
\affiliation{Key Laboratory of Artificial Structures and Quantum Control (Ministry of Education), Department of Physics and 
Astronomy, Shanghai Jiaotong University, 800 Dongchuan Road, Shanghai 200240, China}
\affiliation{Collaborative Innovation Center of Advanced Microstructures, Nanjing, China}

\author{X. L. Lei}
\affiliation{Key Laboratory of Artificial Structures and Quantum Control (Ministry of Education), Department of Physics and 
Astronomy, Shanghai Jiaotong University, 800 Dongchuan Road, Shanghai 200240, China}
\affiliation{Collaborative Innovation Center of Advanced Microstructures, Nanjing, China}

\begin{abstract}

We present a theoretical investigation for the full counting statistics of the Andreev tunneling through a quantum dot (QD) 
embedded between superconducting (SC) and normal leads in the presence of a strong on-site electron-phonon interaction using 
nonequilibrium Green function method. For this purpose, we generalize the dressed tunneling approximation (DTA) recently 
developed in dealing with inelastic tunneling in a normal QD system to the Andreev transport issue. This method takes account of 
vibrational effect in evaluation of electronic tunneling self energy in comparison with other simple approaches and meanwhile 
allows us to derive an explicit analytical formula for the cumulant generating function at the subgap region.
We then analyze the interplay of polaronic and SC proximity effects on the Andreev reflection spectrum, current-voltage 
characteristics, and current fluctuations of the hybrid system. Our main findings include: (1) no phonon side peaks in the linear 
Andreev conductance; (2) a negative differential conductance stemming from the suppressed Andreev reflection spectrum; (3) a 
novel inelastic resonant peak in the differential conductance due to phonon assisted Andreev reflection; (4) enhancement or 
suppression of shot noise for the symmetric or asymmetric tunnel-coupling system respectively.

\end{abstract}

\date{\today}


\maketitle

\section{Introduction}

Modern nanotechnology has facilitated the fabrication of single-electron devices using organic molecules. Since single molecule 
has much smaller mechanical parameters than semiconductor materials, it is very easy to excite the internal vibrational degrees 
of freedom (phonon modes) when electrons are incident upon the single molecule through a tunnel 
junction.\cite{hPark,jPark,Weig,LeRoy}
This has led to experimental observations of a variety of intriguing effects in the transport properties of the single-molecule 
transistors, for instance, the phonon-assisted current steps in the current-voltage characteristics of a variety of individual 
molecule connected to metal electrodes,\cite{hPark,jPark,Weig} the Franck-Condon (FC) blockade in the current steps and negative 
differential conductance (NDC) due to nonequilibrated phonon excitation in the device of a suspended single-wall carbon 
nanotube.\cite{LeRoy}
These experiments have stimulated great interest in theoretical investigations on electronic transport through a quantum dot (QD) 
connected to two electrodes subject to a local strong electron-phonon interaction 
(EPI).\cite{Mitra,Koch,Zazunov,Shen,Frederiksen,Vega,Viljas,Entin,Haule,Flensberg,Rodero,Galperin,Zazunov2,Schmidt,Avriller,Maier
,Dong,Souto}

In experiments, a QD embedded in superconducting (SC) and normal electrodes has attracted intensive studies for over decades, 
since this setup provides one of the most appropriate benchmarks to investigate the interplay between electron correlation and SC 
proximity-induced on-dot pairing effects, for example, in the experimental studies of the Kondo-enhanced Andreev 
transport,\cite{Hofstetter2,Deacon,Franke} Andreev bound states,\cite{Pillet} and Cooper pair splitting.\cite{Hofstetter} Very 
recently, a carbon nanotube QD has been successfully connected to a Nb SC and a normal metal contact and its subgap transport 
properties have been measured, leading to the observation of the phonon assisted resonant Andreev tunneling.\cite{Gramich} 
Actually, this interplay between electron-phonon correlation and SC pairing effects has been theoretically studied in 
literature,\cite{Zhang2,Golez,Zhang,Bocian,Baranski} and predicted recently to be useful in ground-state cooling of a mechanical 
oscillator.\cite{Stadler} 
Based on the nonequilibrium Green function (NGF) method and the Lang-Firsov transformation, two simple decoupling approaches have 
been developed to examine the subgap transport in the polaronic regime: the single-particle approximation 
(SPA)\cite{Zhang2,Bocian,Baranski} and the polaron tunneling approximation (PTA).\cite{Zhang} 

However, the two simple schemes are not able to give the electronic spectral density properly for the N-QD-N system, since both 
of them take no account of the vibrational effect on electronic self-energies due to tunneling in their calculations of the 
electronic GF.\cite{Souto} Besides, the two simple schemes have some other drawbacks. For instance, both of them predicted 
appearances of phonon side peaks in the conductance $G$ with varying the gate voltage $\varepsilon_d$ even in the zero bias 
voltage limit, which is inconsistent with the experiment results.\cite{hPark,Mitra,Entin,Dong} We would like to see in the 
following of this paper that those studies for the N-QD-S system based on the two schemes inherit this unphysical result. 
Moreover, the SPA is not a current conservation approximation. While the PTA is albeit a current conservation approximation, it 
only considers the elastic scattering processes during electron tunneling.\cite{Dong}

Fortunately, an advanced scheme over the SPA and PTA has recently been developed to overcome these drawbacks successfully, in 
which the vibrational effect has been taken into account in the calculation of electronic self-energies.\cite{Dong}  This dressed 
tunneling approximation (DTA) is therefore believed to be able to give the electronic spectral function of a EPI system properly, 
and corrects the pathologies of SPA and PTA in the low-energy and high-energy regimes respectively.\cite{Souto} This scheme also 
predicts the correct transport behavior in the linear regime, no phonon side peaks in the $G$-$\varepsilon_d$ curve. Moreover, 
the DTA fulfills the current conservation condition automatically and includes the inelastic tunneling processes 
naturally.\cite{Dong} The most great advantage of this nonperturbative scheme is that it allows us to obtain an explicit 
analytical expression for the full counting statistics (FCS) of inelastic electron tunneling in the strong EPI system.\cite{Dong}
It is therefore very interesting to examine the interplay of the EPI and SC correlation effects on the FCS of electronic Andreev 
tunneling in the subgap regime using the DTA. It is indeed the purpose of the present paper.

The rest of the paper is organized as follows. In Sec.~II, we present our model of the N-QD-S system and discuss the theoretical 
derivation for the FCS in the subgap transport regime in absence and presence of EPI, respectively. In both cases, the explicit 
analytical expressions of the FCS, current, and zero-frequency shot noise are derived. In Sec.~III, we then present and analyze 
our numerical calculations for the electronic Andreev reflection spectrum, linear and nonlinear Andreev conductances, and shot 
noise, followed by a brief summary in Sec.~IV.

\section{Model and theoretical formulation}

\subsection{Model Hamiltonian}

To investigate the vibrational assisted electron tunneling in the N-QD-S system, we consider the simplest QD model in which a 
single electronic level is coupled to a localized vibrational mode, and is also coupled to a normal metal and a SC lead. The 
model Hamiltonian is
\begin{subequations}
\bq
H = H_{QD}+ H_{N}+H_{S}+H_{TN}+H_{TS}, \label{ham}
\eq
with
\bn
H_{QD} &=& \varepsilon_d \sum_{\sigma} [d_{\sigma}^{\dagger} d_{\sigma}^\pdag + g_{ep} d_{\sigma}^\dagger d_{\sigma}^\pdag 
(a^\dagger + a)] + \omega_0 a^\dagger a, \label{Hmol} \\
H_{N} &=& \sum_{ {\bf k}\sigma} \varepsilon_{N {\bf k}} c_{N {\bf k}\sigma}^\dagger c_{N {\bf k}\sigma}^\pdag, \\
H_{S} &=& \sum_{ {\bf k}\sigma} \varepsilon_{S {\bf k}} c_{S {\bf k}\sigma}^\dagger c_{S {\bf k}\sigma}^\pdag  + \sum_{{\bf k}} 
(\Delta c_{S {\bf k}\downarrow}^\dagger c_{S -{\bf k}\uparrow}^\dagger + {\rm H.c.}), \cr
&& \\
H_{TN} &=& \sum_{{\bf k}\sigma} (\gamma_{N} e^{-i\lambda(t)/2} c_{N {\bf k}\sigma}^\dagger d_\sigma + {\rm H.c.}), \\
H_{TS} &=& \sum_{{\bf k}\sigma} (\gamma_{S} c_{S {\bf k}\sigma}^\dagger d_{\sigma} + {\rm H.c.}).
\en
\end{subequations}
Here, $d^\dagger$ ($d$) denotes the operator of an electron with spin $\sigma$ and energy $\varepsilon_d$ at the QD. $a^\dagger$ 
($a$) is phonon creation (annihilation) operator for the vibrational mode with energy quanta $\omega_0$. $g_{ep}$ is the EPI 
strength.
$c_{\eta{\bf k}}^\dagger$ ($c_{\eta{\bf k}}$) is the creation (annihilation) operator of an electron with spin $\sigma$, momentum 
${\bf k}$, and energy $\varepsilon_{\eta {\bf k}}$ in the normal lead $\eta=N$ or the SC lead $\eta=S$. The SC lead is assumed to 
be described by the BCS Hamiltonian with a SC gap $\Delta$.
$\gamma_{\eta}$ describes the tunnel-coupling matrix element between the QD and lead $\eta$. The corresponding coupling strength 
is defined as $\Gamma_\eta = 2\pi \sum_{k} |\gamma_\eta|^2 \delta(\omega-\varepsilon_{\eta {\bf k}})$, which is assumed to be 
independent of energy in the wide band limit. In order to calculate the FCS, an artificially measuring field $\lambda (t)$ is 
introduced with respect to the normal lead on the Keldysh contour (In this paper, we focus our attention on the tunneling and its 
fluctuation measured in the normal lead only, and we therefore assume that the bias voltage is applied to the normal electrode, 
i.e. $\mu_N=\mu+V$, $\mu_S=\mu$, and we set $\mu=0$ at the equilibrium condition): $\lambda(t)=\lambda_{-} \theta(t) \theta({\cal
T}-t)$ on the forward path and $\lambda(t)=\lambda_{+} \theta(t) \theta({\cal T}-t)$ on the backward path (${\cal T}$ is the 
measuring time during which the counting field is non-zero and the counting field will be set to be opposite constants on the 
forward and backward Keldysh contour as $\lambda_{-}=-\lambda_{+}=\lambda$ in the final derivation).\cite{Gogolin} Throughout we 
will use natural units $e=\hbar=k_{\rm B}=1$.

Since we are interested in the case of strong EPI in this paper, it is convenient to eliminate the EPI term in the Hamiltonian 
Eq.~(\ref{ham}) by applying a nonperturbative Lang-Firsov canonical transformation, i.e. $\widetilde{H}=e^{S} H e^{-S}$ with $S=g 
d^\dagger d (a^\dagger - a)$ ($g=g_{ep}/\omega_0$).\cite{Mahan} The transformed Hamiltonian reads
\begin{subequations}
\bn
\widetilde{H}&=& \widetilde{H}_{QD} + H_{N} + H_{S}+ \widetilde{H}_{TN}+ \widetilde{H}_{TS}, \label{tranH}\\
\widetilde{H}_{QD} &=& \widetilde{\varepsilon}_d \sum_{\sigma}d_\sigma^\dagger d_\sigma + \omega_0 a^\dagger a, \\
\widetilde{H}_{TN} &=& \sum_{{\bf k}\sigma} (\gamma_{N} e^{-i\lambda(t)/2} c_{N {\bf k}\sigma}^\dagger d_\sigma X + {\rm H.c.}), 
\label{tunnelingN}\\
\widetilde{H}_{TS} &=& \sum_{{\bf k}\sigma} (\gamma_{S} c_{S {\bf k}\sigma}^\dagger d_\sigma X + {\rm H.c.}), \label{tunnelingS}
\en
where $\widetilde{\varepsilon}_d = \varepsilon_d - \frac{g_{ep}^2}{\omega_0}$ is the shifted energy level of the QD by the 
polaronic binding energy. To simplify notation we still use $\varepsilon_d$ to denote the shifted level in the following of the 
paper. $\widetilde{d}_\sigma=d_\sigma X$ denotes the new Fermionic operator dressed by the polaronic shift operator $X$,
\bq
X = e^{g (a-a^\dagger)}.
\eq
\end{subequations}

\subsection{Adiabatic Potential for FCS}

The transport problem in the N-QD-S system can be solved with the Keldysh NGF technique in the Nambu space, in which a mixture 
Fermion operator, $\widetilde{\psi}_d= (\widetilde{d}_{\uparrow}, \widetilde{d}_{\downarrow}^\dagger)^T$, has to be introduced to 
describe electronic dynamics involving SC correlation. Accordingly, we must define the contour-ordered GF of the QD, 
$G_d(t,t')=-i\langle T_{\cal C} \widetilde{\psi}_d(t) \widetilde{\psi}_d^\dagger(t') \rangle$ ($T_{\cal C}$ denotes time ordering 
along the Schwinder-Keldysh contour), and the GF of the decoupled lead $\eta$, $g_{\eta {\bf k}}(t,t')=-i\langle T_{\cal C} 
\psi_{\eta {\bf k}}(t) \psi_{\eta {\bf k}}^\dagger(t') \rangle$ with $\psi_{\eta {\bf k}}=(c_{\eta {\bf k}\uparrow}, c_{\eta 
-{\bf k}\downarrow}^\dagger)^T$, in the Nambu representation.

The focus of the present work is the FCS of the subgap Andreev transport in the hybrid tunneling junction, which can be obtained 
from the cumulant generating function (CGF) as a Keldysh partition function
\bq
\chi(\lambda)= \left\langle T_{\cal C} e^{-i \int_{\cal C} (\widetilde{H}_{TN}(t) +\widetilde{H}_{TS}(t))dt} \right\rangle.
\eq
Following the procedure outlined in Ref.~\onlinecite{Gogolin}, the CGF is in turn related to the adiabatic potential
$-i{\cal T } {\cal U}(\lambda)=\ln \chi(\lambda)$, whose derivative in the counting field $\lambda_{-}$ can be expressed 
as\cite{Dong,Gogolin}
\begin{eqnarray}
{\partial {\cal U}(\lambda)\over {\partial \lambda_{-}}} &=& \left\langle {\partial \widetilde{H}_{TN}(t)\over{\partial 
\lambda_{-}}} \right\rangle \nonumber\\
&=& -{i\over 2}\sum_{{\bf k}\sigma} \left\langle \gamma_N e^{-i\lambda_{-}/2} c_{N {\bf k}\sigma}^\dagger(t) 
\widetilde{d}_\sigma(t) - {\rm H.c.} \right\rangle. \cr
&&
\end{eqnarray}
Employing the counting field-dressed GFs defined above, we can easily formulate the derivative of the adiabatic potential in the 
counting field as
\bn
{\partial {\cal U}(\lambda)\over {\partial \lambda_{-}}}&=& \sum_{\bf k}\frac{\gamma_N^2}{2} \int d\omega {\rm Tr}_N \left[ 
\Lambda e^{-i \bar{\lambda}/2} G_{d}^{-+}(\omega,\lambda) g_{N {\bf k}}^{+-}(\omega) \right. \cr
&& \left. - \Lambda^\dagger e^{i\bar{\lambda}/2} g_{N {\bf k}}^{-+}(\omega) G_{d}^{+-}(\omega,\lambda) \right ], \label{adp1}
\en
where
\bq
\Lambda= \left (
\begin{array}{cc}
e^{-i\lambda} & 0 \\
0 & -e^{i\lambda}
\end{array}
\right ),
\eq
and ${\rm Tr}_N[\cdots]$ is the trace over the Nambu space.

\subsection{Noninteracting case}

First we derive the general form of the adiabatic potential ${\cal U}(\lambda)$ for a noninteracting N-QD-S system. In this case, 
what we need is the GF of the QD obtained from a counting field $\lambda$ dressed version of the Dyson equation in the frequency 
space,
\begin{widetext}
\bq
G_d^{-1}(\omega,\lambda)= \left (
\begin{array}{cccc}
\omega- \epsilon_d - \Sigma_{11}^{-+}-\Sigma_{11}^r & \Sigma_{11\lambda}^{-+} & -(\Sigma_{12}^{-+}+ \Sigma_{12}^r) & 
\Sigma_{12\lambda}^{-+}\\
\Sigma_{11\lambda}^{+-} & -(\omega-\epsilon_d+ \Sigma_{11}^{+-} - \Sigma_{11}^r) & \Sigma_{12\lambda}^{+-} & 
-(\Sigma_{12}^{+-}-\Sigma_{12}^{r})\\
-(\Sigma_{21}^{-+}+ \Sigma_{21}^{r}) & \Sigma_{21\lambda}^{-+} & \omega+ \varepsilon_d-\Sigma_{22}^{-+}-\Sigma_{22}^{r} & 
\Sigma_{22\lambda}^{-+} \\
\Sigma_{21\lambda}^{+-} & -(\Sigma_{21}^{+-}-\Sigma_{21}^{r}) & \Sigma_{22\lambda}^{+-} & -(\omega+\varepsilon_d 
+\Sigma_{22}^{+-}-\Sigma_{22}^{r})
\end{array}
\right ), \label{gf0}
\eq
\end{widetext}
where the electronic self-energies $\Sigma_{\alpha\beta\lambda}^{\pm \mp,r}$ ($\alpha,\beta=1,2$ are the Nambu indices) are 
stemming from the tunnel-coupling contributions of the electronic degree of freedom on the QD with both the normal and SC leads, 
$\Sigma_{\alpha\beta\lambda}^{\pm\mp,r}=\Sigma_{N\lambda,\alpha\beta}^{\pm\mp,r}+\Sigma_{S,\alpha\beta}^{\pm\mp,r}$,
\begin{subequations}
\bq
\Sigma_{N\alpha\beta}^r =-i\frac{1}{2}\Gamma_N \delta_{\alpha\beta}, \label{seN}
\eq
\bn
\Sigma_{N\lambda}^{-+}&=& i\Gamma_N \left (
\begin{array}{cc}
f_N(\omega) e^{i\lambda} & 0 \\
0 & [1-f_N(-\omega)] e^{-i\lambda}
\end{array}
\right ) , \\
\Sigma_{N\lambda}^{+-}&=& -i\Gamma_N  \left (
\begin{array}{cc}
[1-f_N(\omega)]e^{-i\lambda} & 0 \\
0 & f_N(-\omega)e^{i\lambda}
\end{array}
\right ),
\en
\bq
\Sigma_{N}^{\pm\mp}=\Sigma_{N\lambda}^{\pm\mp} \mid_{\lambda=0},
\eq
and
\bq
\Sigma_{S}^r = -i \frac{1}{2}\Gamma_S \beta_S(\omega) \left (
\begin{array}{cc}
1 & -\frac{\Delta}{\omega} \\
-\frac{\Delta}{\omega} & 1
\end{array}
\right ) , \label{seS}
\eq
\bq
\Sigma_{S}^{\mp\pm} = \pm i \Gamma_S \Re\beta_S(\omega) \left (
\begin{array}{cc}
1 & -\frac{\Delta}{\omega} \\
-\frac{\Delta}{\omega} & 1
\end{array}
\right ) f_S(\pm \omega),
\eq
with
\bq
\beta_S(\omega)= \frac{|\omega|  \theta(|\omega| - \Delta)}{\sqrt{\omega^2-\Delta^2}} -i \frac{\omega \theta(\Delta- 
|\omega|)}{\sqrt{\Delta^2 - \omega^2}},
\eq
and the Fermi distribution function $f_\eta(\omega)=[e^{(\omega-\mu_\eta)/T}+1]^{-1}$ at lead $\eta$.
\end{subequations}

It should be noticed that we write the GF Eq.~(\ref{gf0}) in the direct product space of the Keldysh and Nambu spaces. This is 
the reason that the GF has a $4\times 4$ matrix form. Simply calculating the derivative of the GF Eq.~(\ref{gf0}) with respect to 
the counting field $\lambda$ gives
\begin{widetext}
\bq
\frac{d}{d\lambda}G_d^{-1}(\omega,\lambda)= \left (
\begin{array}{cccc}
0 & -\Gamma_N f_N(\omega)e^{i\lambda} & 0 & 0\\
-\Gamma_N[1-f_N(\omega)]e^{-i\lambda} & 0 & 0 & 0\\
0 & 0 & 0 & \Gamma_N [1-f_N(-\omega)]e^{-i\lambda} \\
0 & 0 & \Gamma_N f_N(-\omega)e^{i\lambda} & 0
\end{array}
\right ). \label{dgf0}
\eq
\end{widetext}
Then we apply Jacobi's formula for the derivative of the determinant of a matrix
\bq
\frac{d}{d\lambda}\det(G_d^{-1}(\lambda))={\rm Tr} \left [{\rm adj}(G_d^{-1}(\lambda)) \frac{d}{d\lambda} G_d^{-1}(\lambda) 
\right],
\eq
to yield
\bn
\frac{1}{\det(G_d^{-1}(\omega,\lambda))} \frac{d}{d\lambda}\det(G_d^{-1}(\omega,\lambda)) &=& \cr
&& \hspace{-3cm} {\rm Tr} \left [ G_d(\omega,\lambda) \frac{d}{d\lambda}G_d^{-1}(\omega,\lambda) \right ],
\en
where ${\rm Tr}[\cdots]$ is the trace over the direct product space of the Keldysh and Nambu spaces, not over the Nambu space 
only as in Eq.~(\ref{adp1}). Utilizing Eq.~(\ref{dgf0}) to perform the trace calculation of right hand side of the last equation, 
we can find that it is exactly equal to the integrand in the right hand side of Eq.~(\ref{adp1}). Then after integrating with 
respect to the counting field, we can obtain the general formula for the adiabatic potential
\bq
{\cal U}(\lambda)=i\int \frac{d\omega}{2\pi} \ln \left [ \frac{\det (G_d^{-1}(\omega,\lambda))}{\det (G_d^{-1}(\omega,0))} \right 
]. \label{adp2}
\eq
This is a direct generalization of the Fredholm determinant of the CGF to the noninteracting system involving SC electrode.

In order to investigate the current and noise, we now have to calculate the determinants inside the logarithm in 
Eq.~(\ref{adp2}). It is not quite an easy task. However, considering that we will focus our attention on the electron tunneling 
only in the subgap region at temperature and bias voltage well below the SC gap, $T\ll \Delta$ and $V\ll \Delta$, in the present 
investigation, we can take $\Sigma_{S}^{\pm\mp}=0$ in our following derivation. To the end, we can obtain the general form for 
the adiabatic potential for the subgap Andreev tunneling
\bn
{\cal U}_{A}(\lambda) &=& i \int \frac{d\omega}{2\pi} \ln \left \{ 1+ T_A(\omega) [ (e^{i2\lambda}-1)f_N f_{-N} \right.\cr
&& \left. + (e^{-i2\lambda}-1) (1-f_N) (1-f_{-N}) ] \right \}, \label{adp3}
\en
with the shorthand notations $f_N=f_N(\omega)$ and $f_{-N}=f_N(-\omega)$, and the Andreev reflection probability 
$T_A(\omega)=\Gamma_N^2 |G_{12}^r(\omega)|^2$. It is found that the subgap FCS remains binomial form but with a double-charge 
transfer. From Eq.~(\ref{adp3}), we can distinguish the elementary processes of electronic tunneling in the subgap region: (i) 
two opposite-spin electrons having energies equally higher and lower the chemical potential are respectively transmitted from the 
normal lead to the QD, and eventually enter into the SC lead to form a Cooper pair. The probability of this process is 
$P_+=T_A(\omega) f_N f_{-N}$. In literature, this is considered as Andreev reflection that an electron incident from the normal 
lead with energy $\omega$ tunnels into the QD, and subsequently picks up an opposite-spin electron with energy $-\omega$ to 
create a Cooper pair into the SC lead and leads to a hole propagating back to the normal lead; (ii) the reverse process with a 
probability $P_-=T_A(\omega) (1-f_N)(1-f_{-N})$; (iii) no transmission happens with a probability $P_0=1-P_+-P_-$.

From Eq.~(\ref{adp3}) we can evaluate the current and noise for the Andreev tunneling, respectively, as
\bn
I_A &=& -\frac{\partial U_A}{\partial\lambda}{\bigg |}_{\lambda=0} = -2\int \frac{d\omega}{2\pi} T_A(\omega) \left ( 1-f_N- 
f_{-N} \right),\cr
&& \\
S_A &=& i \frac{\partial^2 U_A}{\partial\lambda^2}{\bigg |}_{\lambda=0} = 4 \int \frac{d\omega}{2\pi}\left [ T_A(\omega) \left ( 
1-f_N- f_{-N} \right. \right. \cr
&& \left. + 2f_Nf_{-N} \right ) - T_A^2(\omega) \left ( 1-f_N - f_{-N})^2 \right ]. \label{sano}
\en
The differential Andreev conductance $g_A$ is obtained by differentiating $I_A$ with respect $V$ as
\bq
g_A = \frac{2e^2}{h} \frac{1}{T}\int d\omega T_A(\omega) [f_-(1-f_-) + f_+(1-f_+)], \label{gPTA}
\eq
with $f_{\pm}=[e^{(\omega \pm V)/T}-1]^{-1}$. The noise $S_A$ can be rewritten as the sum of the equilibrium noise (thermal 
noise) $S_{th}$ and the nonequilibrium noise (shot noise) $S_{sh}$,
\bn
S_A &=& \frac{4e^2}{h} \int d\omega \{ T_A(\omega) [f_-(1-f_-) + f_+(1-f_+)] \cr
&& + T_A(\omega) [1-T_A(\omega)] (f_--f_+)^2 \}.
\en
It can therefore be realized that the generalized nonequilibrium Nyquist-Johnson (NNJ) relation becomes
\bq
g_A= \frac{1}{2T} (S_A - S_{sh}),
\eq
for the two-particle correlation tunneling in the subgap region (double-charge transfer is relevant), in comparison with the NNJ 
relation for the usual single-particle tunneling in the normal system, $g= \frac{1}{4T} (S - S_{sh})$.

\subsection{Electron-phonon interaction system}

Now we generalize our above discussion for the FCS of the Andreev tunneling to the strong EPI case. For this purpose, we still 
need to evaluate the counting field $\lambda$ dressed Dyson equation for the GF of the QD. In the limit of a weak 
tunnel-couplings, $\Gamma_{\eta} \ll \omega_0$, the GF $G_d(t,t')$ in the Nambu representation can be decomposed into a product 
of a pure electronic part $G_c(t,t')=-i\langle T_{\cal C} \psi_d(t) \psi_d^\dagger(t') \rangle$ and a phononic part 
$K(t,t')=\langle T_{\cal C} {\cal K}(t) {\cal K}^\dagger(t') \rangle$ (the Born-Oppenheimer adiabatic 
approximation),\cite{Flensberg,Galperin,Maier,Dong,Souto}
\bq
G_{d\alpha\beta}(t,t') \approx G_{c\alpha\beta}(t,t') K_{\alpha\beta}(t,t'), \label{deGF}
\eq
with $\psi_d= (d_{\uparrow}, d_{\downarrow}^\dagger)^T$ and ${\cal K}=(X, X^\dagger)^T$.

Another approximation employed in this paper is that the vibrational mode is assumed to be always at the equilibrium state, which 
leads to\cite{Mahan}
\begin{subequations}
\bn
K_{11}^{\mp\pm}(t,t')&=& K_{22}^{\mp\pm}(t,t')= \exp \{ -g^2 \left [ n_B(1-e^{\mp i\omega_0 \tau}) \right.\cr
&& \left. + (n_B+1) (1-e^{\pm i\omega_0 \tau}) \right ]\}, \\
K_{12}^{\mp\pm}(t,t') &=& K_{21}^{\mp\pm}(t,t') = \exp \{ -g^2 \left [ n_B(1+e^{\mp i\omega_0 \tau}) \right. \cr
&& \left. + (n_B+1) (1+e^{\pm i\omega_0 \tau}) \right ]\},
\en
\end{subequations}
with the Bose distribution $n_B=(e^{\omega_0/T}-1)^{-1}$ at the temperature $T$ and $\tau=t-t'$. Using the identity $\exp(z\cos 
\theta)=\sum_{n=-\infty}^{\infty} I_n(z) \exp(in\theta)$, these correlation functions can be expanded in a power series in $\exp 
(\pm i\omega_0 \tau)$
\begin{subequations}
\bn
K_{11}^{\mp\pm}(t,t') &=& K_{22}^{\mp \pm}(t,t') = \sum_{n=-\infty}^{\infty} w_n e^{\pm in\omega_0 \tau},\\
K_{12}^{\mp\pm}(t,t') &=& K_{21}^{\mp\pm}(t,t') = \sum_{n=-\infty}^{\infty} (-1)^n w_n e^{\pm in\omega_0 \tau}, \cr
&&
\en
\end{subequations}
with
\bq
w_n=e^{-g^2(2n_B+1)} e^{n\omega_0/2T} I_n(2g^2\sqrt{n_B(n_B+1)}),
\eq
where $I_n(x)$ is the $n$th Bessel function of complex argument. Therefore, in the Fourier domain the dressed GF 
$G_d(\omega,\lambda)$ of the QD can be expressed in terms of the pure electronic GF $G_c(\omega,\lambda)$, which are all 
dependent on the counting field $\lambda$, as
\begin{subequations}
\bn
G_{d11(22)}^{\mp\pm}(\omega,\lambda) &=& \sum_{n=-\infty}^{\infty} w_n G_{c11(22)}^{\mp\pm}(\omega\pm n\omega_0,\lambda), \\
G_{d12(21)}^{\mp\pm}(\omega,\lambda) &=& \sum_{n=-\infty}^{\infty} (-1)^n w_n G_{c12(21)}^{\mp\pm}(\omega\pm 
n\omega_0,\lambda).\cr
&&
\en
\end{subequations}

In what follows, we derive the counting field dressed Dyson equation of the pure electronic GF of the QD, 
$G_{c}(\omega,\lambda)$, based on the transformed Hamiltonian $\widetilde{H}$ Eq.~(\ref{tranH}). There are two ways in literature 
to deal with the exponential operators $X$ and $X^\dagger$ in the transformed tunneling Hamiltonian Eqs.~(\ref{tunnelingN}) and 
(\ref{tunnelingS}).\cite{Dong,Souto} One is to simply ignore the polaronic operators and/or replace them with their respective 
expectation values, $\langle X \rangle$ and $\langle X^\dagger \rangle$, in deriving the equation of motion (EOM) of 
$G_c(\omega)$. Under this approximation, the electronic self-energy is stemming only from the tunnel-coupling between the QD and 
electrodes and no polaronic effect is considered. This approximation is believed to be valid only if the Fermi sea effect of 
electrodes can be neglected in the limit of extremely weak tunnel-coupling, and is consequently called single-particle 
approximation.\cite{Flensberg} The main drawback of the SPA is that it does not obey current conservation condition and 
underestimates the electronic spectral density in the case of a N-QD-N system at low frequencies. To overcome these 
disadvantages, an improved procedure has to be proposed to correctly describe the polaronic effect in tunneling processes, in 
which the Born-Oppenheimer decoupling approximation is invoked once again in the EOM of $G_c(\omega)$ to disentangle the 
polaronic operator and operators of reservoirs yielding vibrational dressed self-energies of electronic tunneling,\cite{Dong}
\bq
\Sigma_{c\alpha\beta\lambda}^{\mp\pm}(t,t')=\Sigma_{\alpha\beta\lambda}^{\mp\pm}(t,t')K_{\beta\alpha}^{\pm\mp}(t',t).
\eq
In this end, we obtain the counting field dressed Dyson equation for $G_c(\omega,\lambda)$ in frequency domain,
\begin{widetext}
\bq
G_c^{-1}(\omega,\lambda)= \left (
\begin{array}{cccc}
\omega- \varepsilon_d - \Sigma_{c11}^{-+}-\Sigma_{c11}^r & \Sigma_{c11\lambda}^{-+} & -(\Sigma_{c12}^{-+}+ \Sigma_{c12}^r) & 
\Sigma_{c12\lambda}^{-+}\\
\Sigma_{c11\lambda}^{+-} & -(\omega-\varepsilon_d+ \Sigma_{c11}^{+-} - \Sigma_{c11}^r) & \Sigma_{c12\lambda}^{+-} & 
-(\Sigma_{c12}^{+-}-\Sigma_{c12}^{r})\\
-(\Sigma_{c21}^{-+}+ \Sigma_{c21}^{r}) & \Sigma_{c21\lambda}^{-+} & \omega+ \varepsilon_d-\Sigma_{c22}^{-+}-\Sigma_{c22}^{r} & 
\Sigma_{c22\lambda}^{-+} \\
\Sigma_{c21\lambda}^{+-} & -(\Sigma_{c21}^{+-}-\Sigma_{c21}^{r}) & \Sigma_{c22\lambda}^{+-} & -(\omega+\varepsilon_d 
+\Sigma_{c22}^{+-}-\Sigma_{c22}^{r})
\end{array}
\right ), \label{gfc}
\eq
\end{widetext}
with ($\alpha\neq \beta$)
\begin{subequations}
\bn
\Sigma_{c\alpha\alpha\lambda}^{\mp\pm}(\omega) &=& \sum_{n=-\infty}^{\infty} w_n \Sigma_{\alpha\alpha\lambda}^{\mp\pm}(\omega\pm 
n\omega_0),\\
\Sigma_{c\alpha\beta\lambda}^{\mp\pm}(\omega) &=& \sum_{n=-\infty}^{\infty} (-1)^n w_n 
\Sigma_{\alpha\beta\lambda}^{\mp\pm}(\omega\pm n\omega_0), \\
\Sigma_{c\alpha\beta}^{\pm\mp}(\omega) &=& \Sigma_{c\alpha\beta\lambda}^{\pm\mp}(\omega) \mid_{\lambda=0}.
\en
\end{subequations}
Moreover, the retarded self-energy in time domain can be defined in the usual way from the lesser and greater counterparts, 
$\Sigma_{c\alpha\beta}^r(\tau)=\theta(\tau) [ \Sigma_{c\alpha\beta}^{+-}(\tau) - \Sigma_{c\alpha\beta}^{-+}(\tau)]$, and thus its 
expression in frequency domain is
\bn
\Sigma_{c\alpha\beta}^r(\omega) &=& i\int \frac{d\omega'}{2\pi} {\cal P} \frac{\Sigma_{c\alpha\beta}^{+-}(\omega')- 
\Sigma_{c\alpha\beta}^{-+}(\omega')}{\omega-\omega'} \cr
&& + \frac{1}{2} \left [ \Sigma_{c\alpha\beta}^{+-}(\omega)- \Sigma_{c\alpha\beta}^{-+}(\omega) \right ] . \label{selfenergy}
\en
Here ${\cal P}$ means principal value integral.
It is noticed that the resultant self-energies $\Sigma_{c\alpha\beta}^r(\omega)$ of electronic tunneling are highly dependent on 
the applied bias voltage. While in the SPA these retarded self-energies are assumed to be equal to those in a noninteracting 
QD-lead system, Eqs.~(\ref{seN}) and (\ref{seS}), which are irrelevant to the bias voltage. We will find in the following that 
this bias voltage dependence of self-energies has profound effect on the tunneling current and its fluctuation in the subgap 
region. It is clear that this voltage dependence is stemming from the polaronic effect, since the phononic propagator is included 
in calculation of the self-energies. That is why this approximation is named as dressed tunneling approximation.\cite{Souto} Two 
advantages of the DTA are that it satisfies the current conservation condition and gives correct spectral density at both low and 
high frequencies in a N-QD-N system.\cite{Dong,Souto}

Another advantage of the DTA is that, when we apply this approximation to investigate the FCS of vibronic assisted tunneling in 
the hybrid N-QD-S system, the general formula for the adiabatic potential of a noninteracting QD system, Eq.~(\ref{adp2}), is 
still applicable as long as the GF $G_d$ is replaced by $G_c$,
\bq
{\cal U}(\lambda)=i\int \frac{d\omega}{2\pi} \ln \left [ \frac{\det (G_c^{-1}(\omega,\lambda))}{\det (G_c^{-1}(\omega,0))} \right
]. \label{adp4}
\eq
Focusing on the subgap tunneling, we can write the subgap adiabatic potential, after lengthy calculations, in an explicit 
expression,
\bn
{\cal U}_{A}(\lambda) &=& i \int \frac{d\omega}{2\pi} \ln \left \{ 1+ \sum_{nm} w_n w_m T_{A}(\omega) \right. \cr
&& \times [ (e^{i2\lambda}-1) f_{N+n} f_{-N+m} + (e^{-i2\lambda}-1) \cr
&& \left. \times (1-f_{N-n}) (1-f_{-N-m}) ] \right \}, \label{adp5}
\en
with $f_{N\pm n}=f_N(\omega \pm n\omega_0)$ and $f_{-N\pm n}=f_N(-\omega \pm n\omega_0)$, and the vibrational dressed Andreev 
reflection probability
\bq
T_{A}(\omega)=\Gamma_N^2 |G_{c12}^r(\omega)|^2. \label{ar}
\eq

This explicit analytical expression Eq.~(\ref{adp5}) provides us a clear physical picture for an elementary phonon-assisted 
Andreev reflection process: an electron incident from the  normal lead with energy $\omega$ absorbs (or emits) $n>0$ ($n<0$) 
phonon and tunnels into the QD, then picks up an electron with energy $-\omega$ to create a Cooper pair, and simultaneously emits 
(or absorbs) $m>0$ ($m<0$) phonon, finally becomes a hole to come back to the normal lead.

Moreover, this expression Eq.~(\ref{adp5}) allows us to derive the formulae of phonon-assisted Andreev tunneling current and its 
shot noise as follows:
\bn
I_A &=& - {\partial U_A(\lambda)\over {\partial (\lambda)}} {\bigg |}_{\lambda=0}
= -2 \int d\omega  \sum_{nm} w_n w_m T_{A}(\omega)  \cr
&& \times \left [ (1-f_{N-n}) (1-f_{-N-m}) - f_{N+n} f_{-N+m} \right ], \cr
&& \label{current}
\en
\begin{widetext}
\begin{eqnarray}
S_A = i {\partial^2 U_A(\lambda) \over {\partial \lambda^2}} {\bigg |}_{\lambda=0} &=& 4 \int d\omega \left ( \sum_{nm} w_n w_m 
T_{A}(\omega) [ (1-f_{N-n}) (1-f_{-N-m}) + f_{N+n} f_{-N+m}) ] \right. \cr
&& \left. - \left \{ \sum_{nm} w_n w_m T_{A}(\omega) [ (1-f_{N-n}) (1-f_{-N-m}) - f_{N+n} f_{-N+m} ] \right \}^2 \right ). 
\label{automf}
\en
\end{widetext}
From Eq.~(\ref{current}) the current can be separated as two contributions of elastic and inelastic parts, $I_A=I_{el}+I_{in}$, 
where the elastic current is
\bq
I_{el} = -2 \int d\omega \, w_0^2 \, T_A(\omega) \left ( 1- f_{N}  - f_{-N} \right ). \label{currentel}
\eq

\section{Results and Discussions}

In this section, we perform numerical investigation of vibronic effect on Andreev reflection properties of a hybrid N-QD-S system 
at the subgap region, by calculating the tunneling current and its shot noise based on Eqs.~(\ref{current}) and (\ref{automf}). 
In numerical calculations, we set the SC gap $\Delta=1$ as the energy unit and choose the Fermi levels of the two leads as the 
reference of energy $\mu_{N}=\mu_{S}=\mu=0$ at equilibrium. We also set the phonon energy $\omega_0=0.2\Delta$, since numerical 
fits of the experimentally measured data for the $I$-$V$ curves predict the vibrational frequency ranging from a few $100\mu$eV 
to a few meV and a strong EPI $g\leq 1$ for suspended carbon nanotubes. In the following calculations, we will choose a symmetric 
hybrid system in the tunneling rates, $\Gamma_S=\Gamma_N=0.1\Delta$, as an example, and also a strongly asymmetric system with 
$\Gamma_S=10\Gamma_N$ as well for comparison with the results of experimental measurements.

\subsection{Self energy and Andreev reflection spectrum}

\begin{figure}[htb]
\includegraphics[height=2.5cm,width=4.3cm]{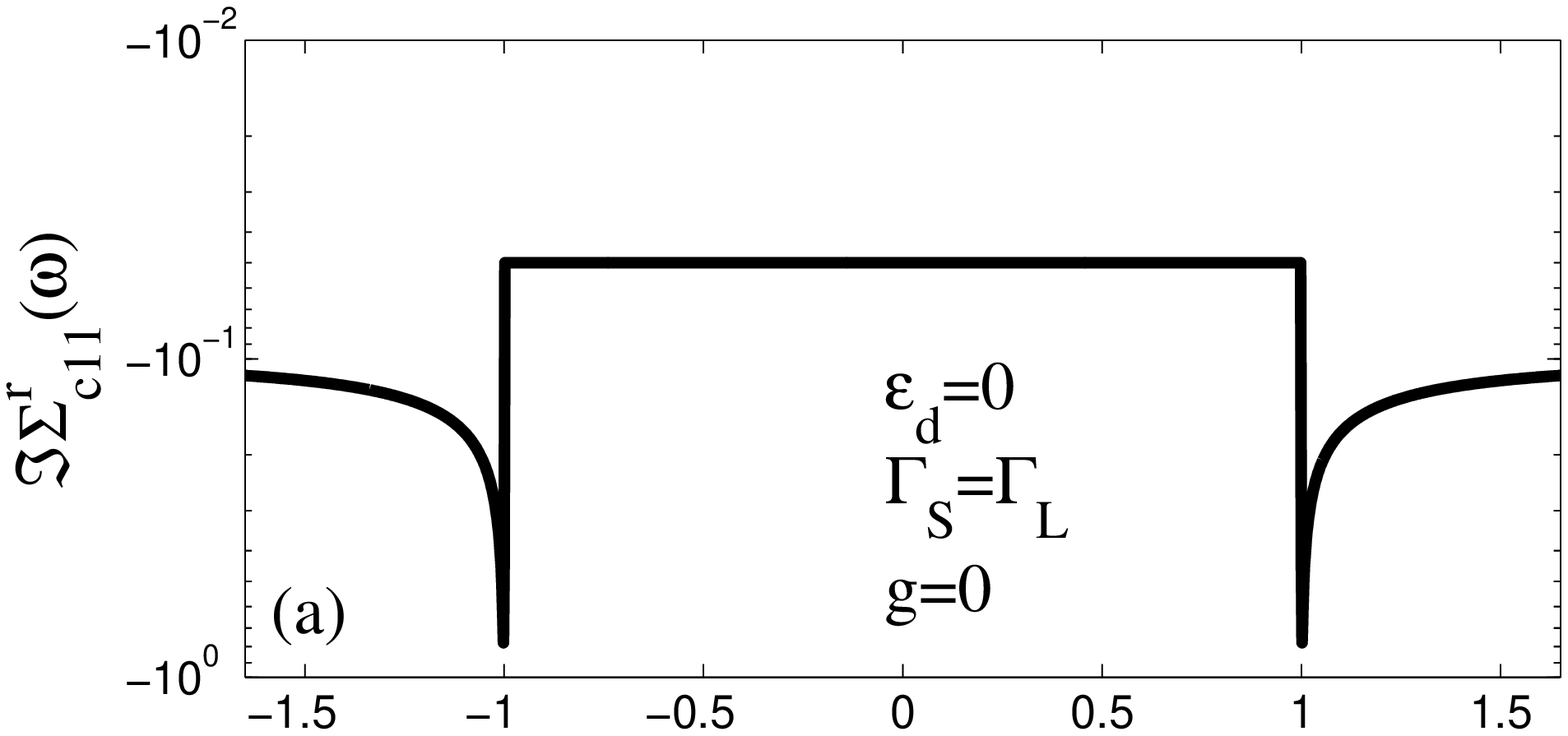}
\includegraphics[height=2.5cm,width=4cm]{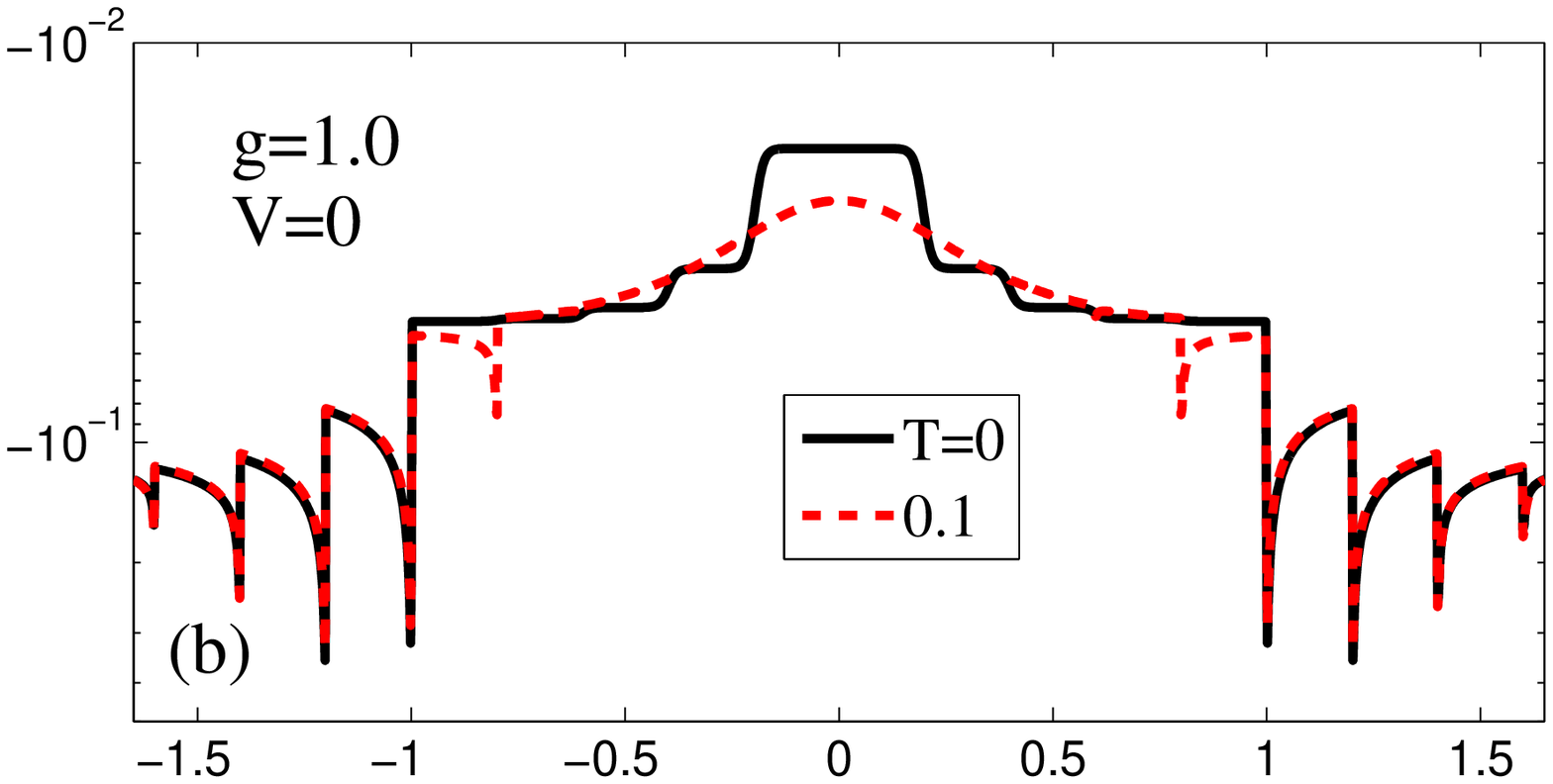}
\vspace{1mm}

\includegraphics[height=2.5cm,width=4.3cm]{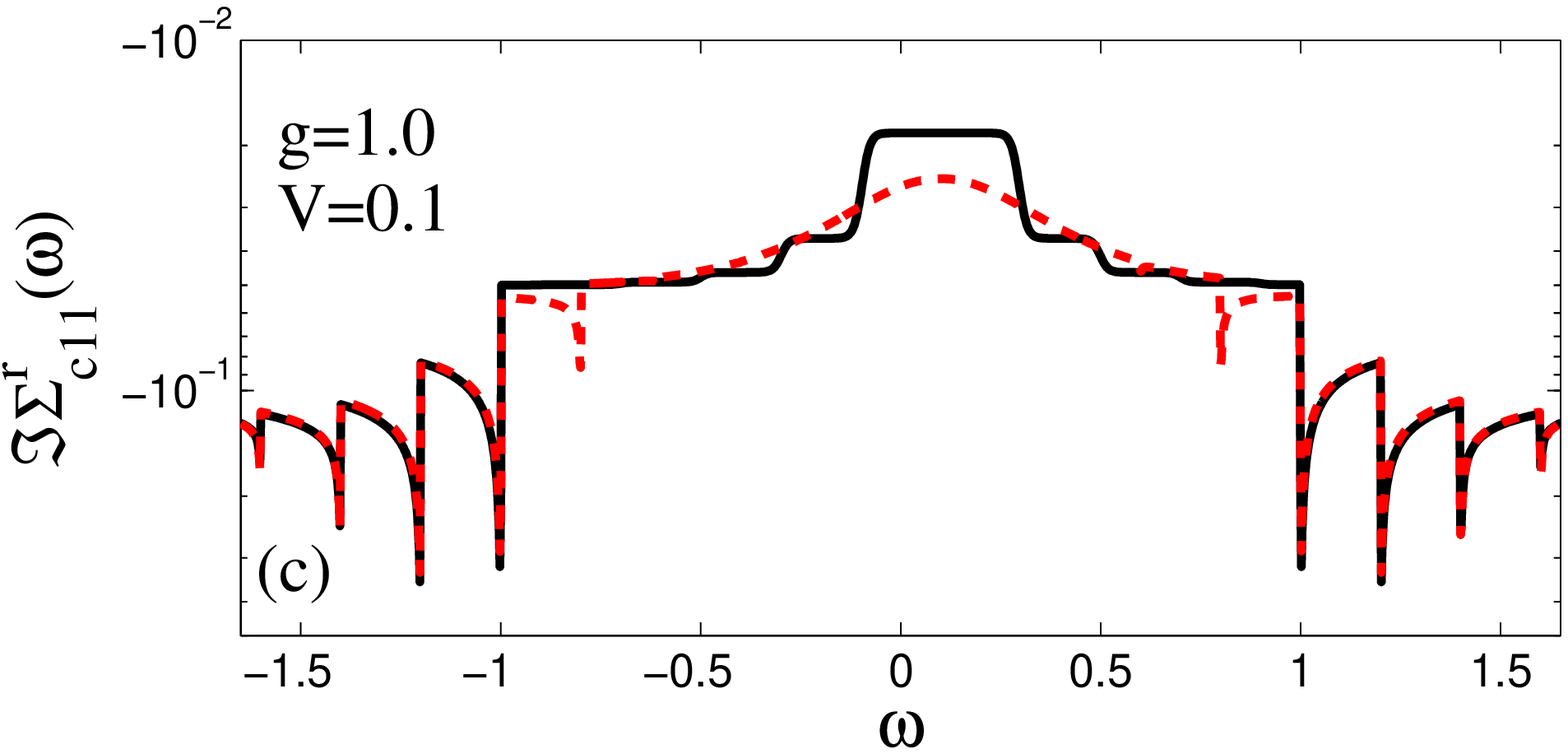}
\includegraphics[height=2.5cm,width=4cm]{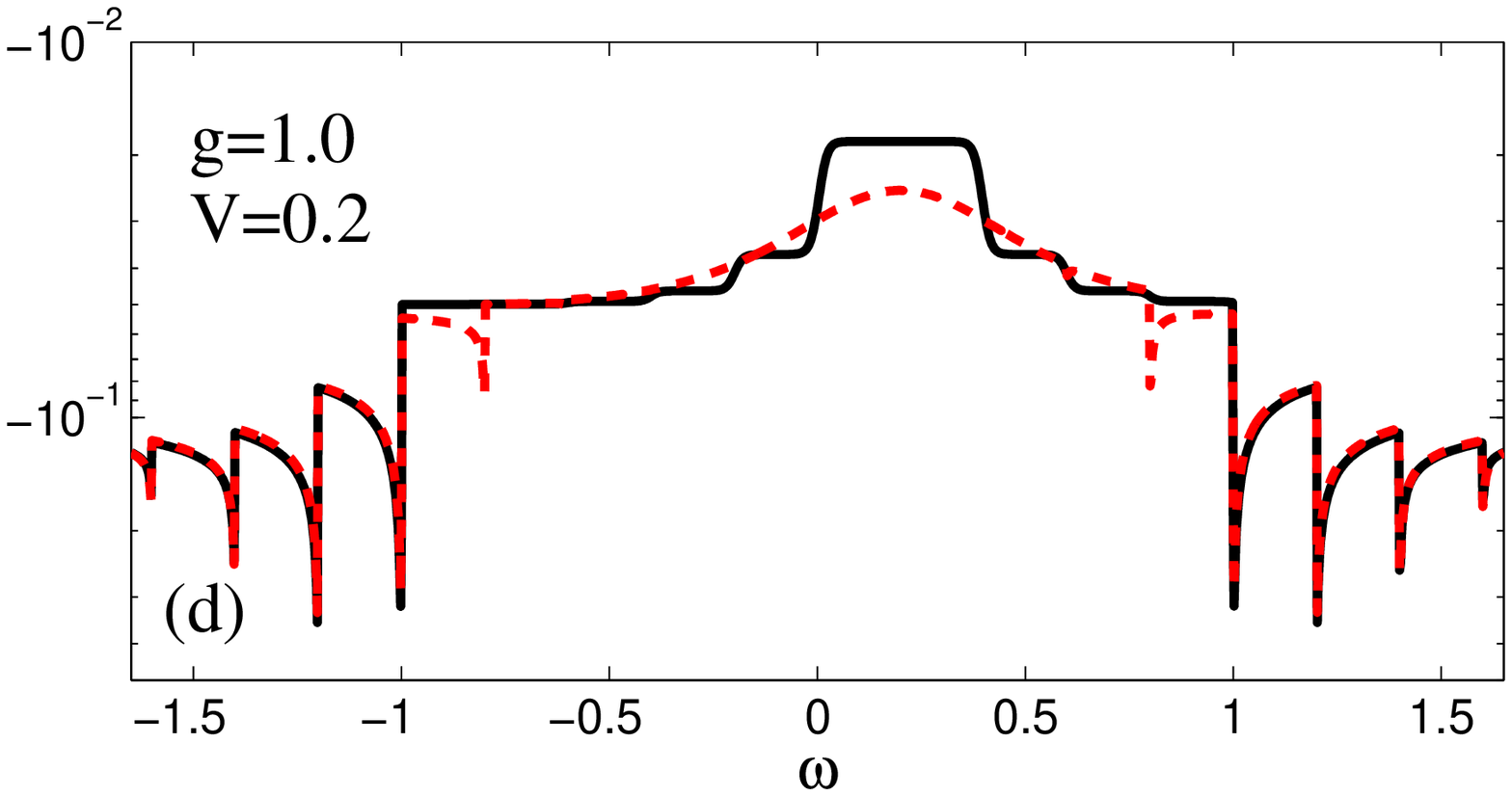}
\caption{(Colour online) The imaginary part of the first diagonal element of the vibrational dressed retarded self-energy, 
$\Sigma_{c11}^r(\omega)$, are plotted for different bias-voltages, $V=0$ (b), $0.1$ (c), and $0.2$ (d), respectively, at 
different temperatures, $T=0$ and $0.1$. The parameters used for calculation are taken as: $\Gamma_N=\Gamma_S=0.1$, $g=1.0$. For 
comparison, the corresponding result for the system without EPI is plotted in (a).}
\label{figse1}
\end{figure}

At first, we examine the dependence of the tunneling-induced electronic self-energy, Eq.~(\ref{selfenergy}), on the bias voltage 
and the temperature due to EPI under DTA. In Fig.~\ref{figse1}, we plot the imaginary part of the first diagonal element of the 
vibrational dressed retarded self-energy, $\Sigma_{c11}^r(\omega)$, for the symmetric system with $g=0$ (a) and $g=1$ (b-d) at 
different bias voltages and temperatures. Without EPI, this quantity shows an obvious discontinuous at $\omega=\pm \Delta$ due to 
SC gap, and no voltage and temperature dependence [see Eqs.~(\ref{seN}) and (\ref{seS})].
In the presence of EPI, we find from Fig.~\ref{figse1}(b) that it not only has the SC gap at $\omega=\pm \Delta$, but also 
develops vibronic replicas of the SC gap edges separated by integer multiples of phonon frequency $\omega_0$ at the regions of 
$|\omega|>\Delta$ at low temperature; meanwhile at the subgap region, $|\omega|<\Delta$, it exhibits explicit stepwise structures 
whose widths are $2\omega_0$ or $\omega_0$, and heights are controlled by the Franck-Condon (FC) factors.
These peaks and steps are all related to the opening of the inelastic transport channels, i.e. phonon assisted normal tunneling 
and/or Andreev reflection. Applying bias voltage just moves the steps in the subgap region towards the positive direction of 
frequency. At the corresponding points of peaks and steps, the real part of the self-energy shows peaks with singularities due to 
Kramers-Kronig relations (not shown here). Increasing temperature has two effects, smoothing the stepwise structures and 
developing additional SC edges within the subgap region.

\begin{figure}[htb]
\includegraphics[height=5cm,width=8.5cm]{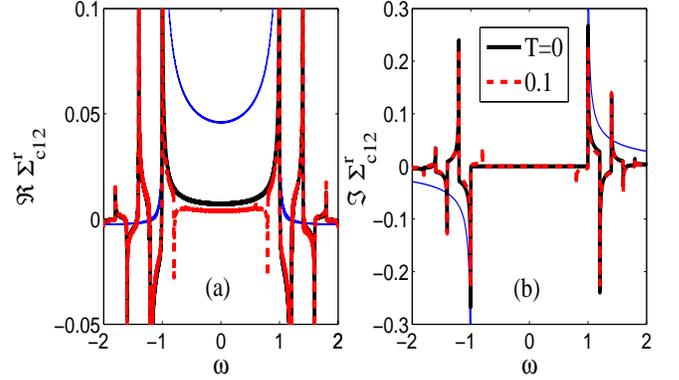}
\caption{(Colour online) The real (a) and imaginary (b) parts of the nondiagonal element of the vibrational dressed retarded 
self-energy, $\Sigma_{c12}^r(\omega)$, at $V=0$ and different temperatures, $T=0$ and $0.1$. The other parameters are the same as 
those in Fig.~\ref{figse1}. The thin-blue lines denote results for the system with $g=0$.}
\label{figse2}
\end{figure}

Figure \ref{figse2} shows the nondiagonal element of the vibrational dressed retarded self-energy, $\Sigma_{c12}^r(\omega)$, at 
$V=0$. Its real part is found to be nonzero at the subgap region for the noninteracting system, which is ascribed to the SC 
proximity effect. According to Eq.~(\ref{seS}), it is equal to $\frac{1}{2} \frac{\Gamma_S\Delta}{\sqrt{\Delta^2-\omega^2}}$, 
while its imaginary part is equal to $0$. In the presence of EPI, the real part at the subgap region is suppressed due to the FC 
blockade.

We now analyze the Andreev reflection spectrum $T_A(\omega)$, since it is an underlying quantity to describe the nonequilibrium 
transport through the hybrid system in the subgap regime. We will compare the DTA results for $T_A(\omega)$, Eq.~(\ref{ar}), with 
those from other approximations. The Andreev reflction spectrum of the SPA can be obtained by simply replacing the dressed self 
energies Eq.~(\ref{selfenergy}) with the bare ones, Eqs.~(\ref{seN}) and (\ref{seS}), in calculating the pure electronic retarded 
GF $G_{c12}^r(\omega)$ in the formula Eq.~(\ref{ar}) of Andreev reflection probability.
In addition to SPA and DTA, another approximation namely as the polaron tunneling approximation is usually utilized to 
investigate the phonon assisted nonequlibrium transport through a normal-QD system perturbatively in strong coupling regime in 
literature.\cite{Maier} The underlying physical essence of PTA is based on the atomic limit, within which the dot GF can be write 
down as the isolated QD after Lang-Firsov transformation and then use Dyson equation to couple the electrodes.
It is nevertheless believed to be overestimated the electronic spectral density at high frequencies and predicts appearances of 
phonon side peaks in the conductance with varying energy level of QD in linear transport regime, which is inconsistent with 
previous experiment measurements.\cite{hPark,LeRoy} It is reported that the two drawbacks can be cured by the more rigorous DTA 
method.\cite{Dong,Souto}

\begin{figure}[htb]
\includegraphics[height=4cm,width=8cm]{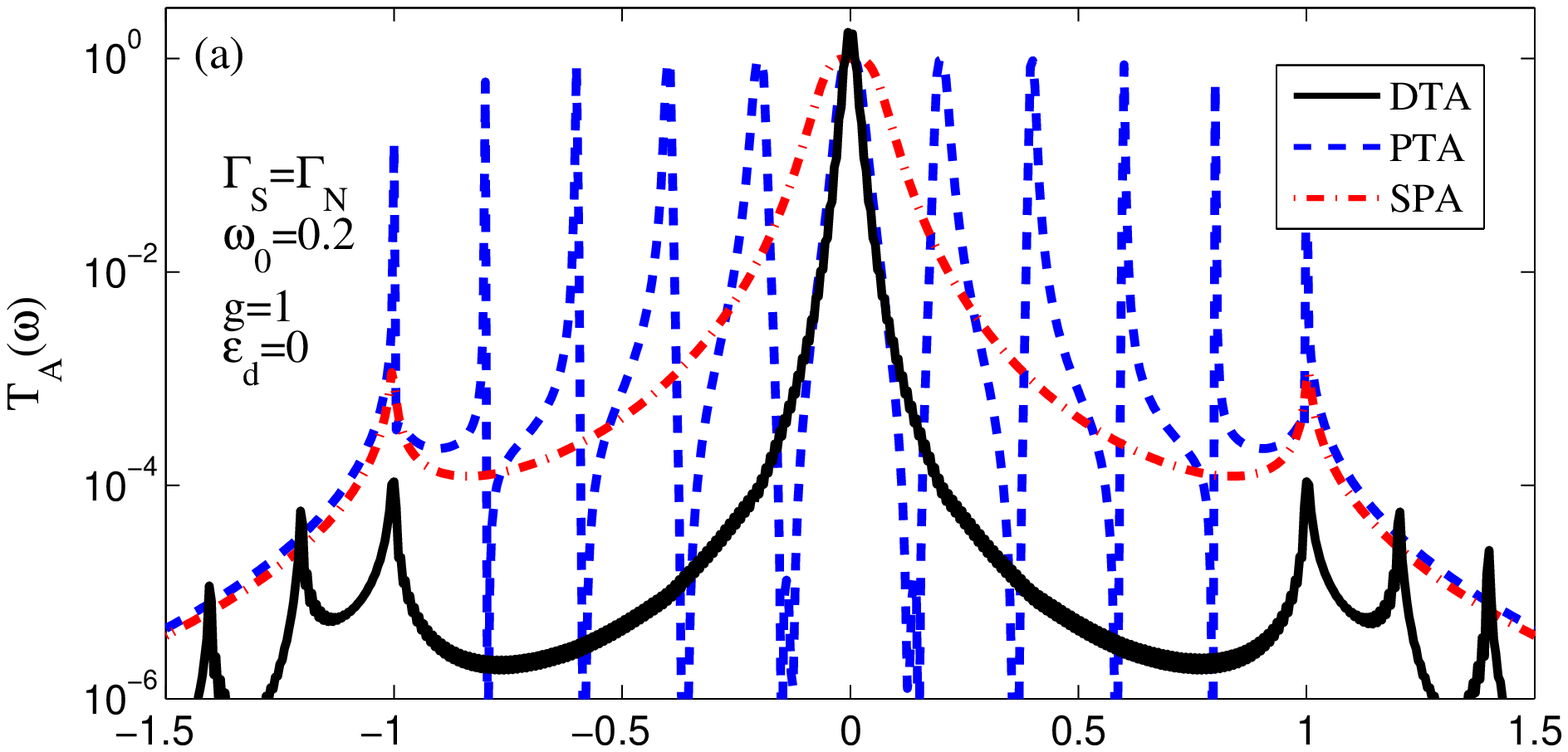}
\vspace{3mm}

\includegraphics[height=4cm,width=8cm]{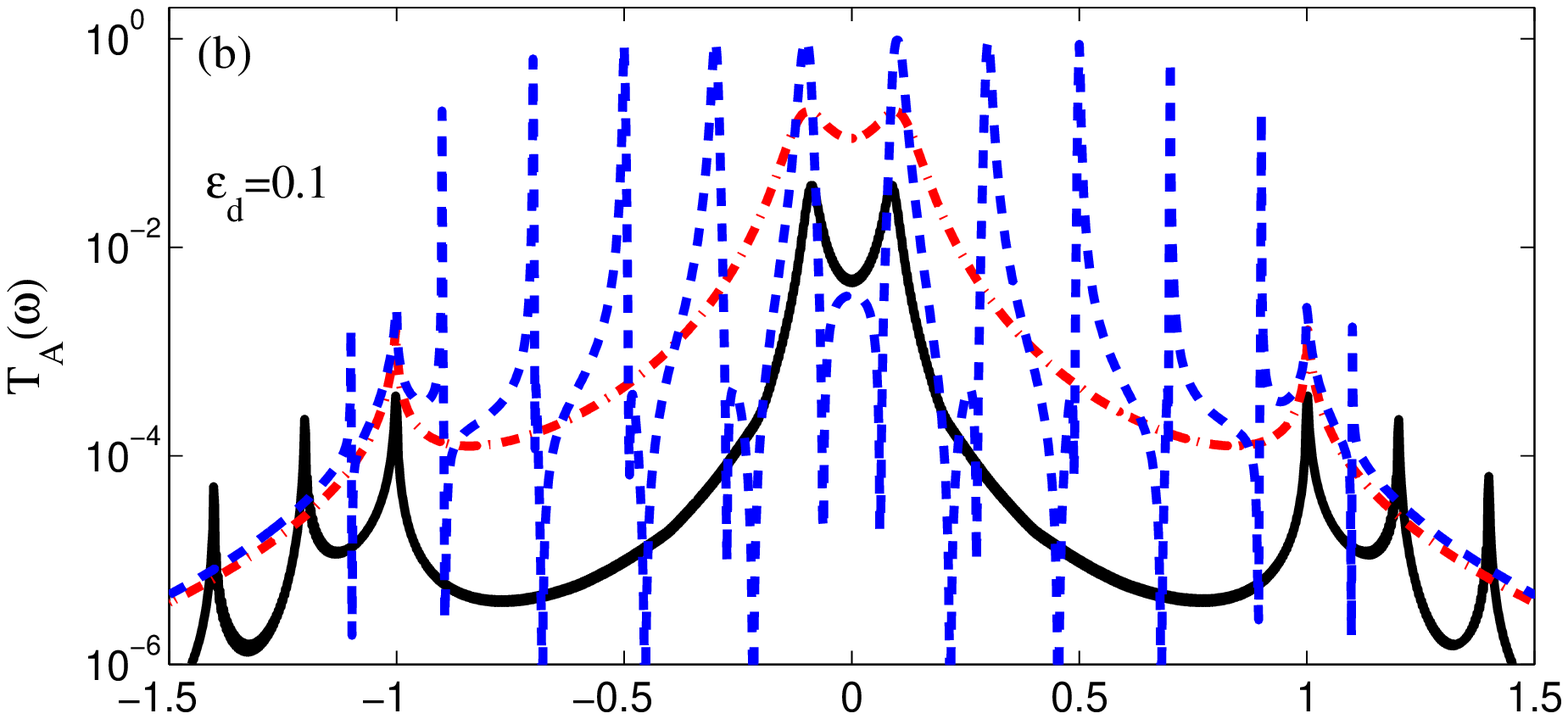}
\vspace{3mm}

\includegraphics[height=4cm,width=8cm]{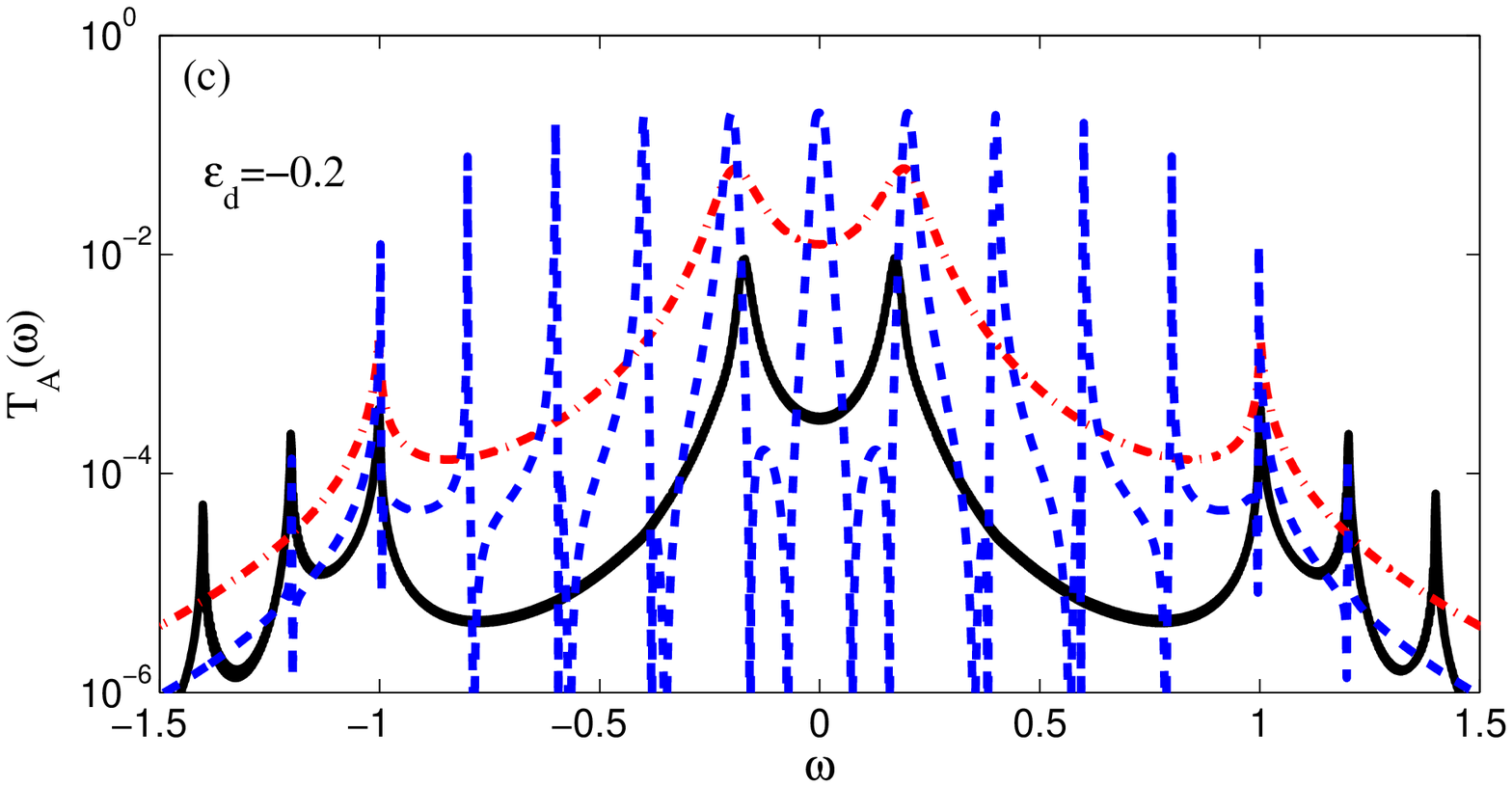}
\caption{(Colour online) The Andreev reflection spectrum for the symmetric N-QD-S system $\Gamma_S=\Gamma_N$ with (a) 
$\varepsilon_d=0$, (b) $0.1$, and (c) $-0.2$ at zero bias voltage and zero temperature. The calculated results correspond to the 
different approximations: DTA (solid-black line), PTA (dashed-blue line), and SPA (dashed-dotted-red line).}
\label{figars}
\end{figure}

Very recently, the PTA has been extended to study phonon assisted Andreev tunneling in the hybrid N-QD-S system, in which the 
bare retarded GF of the QD has the form\cite{Zhang}
\begin{subequations}
\bn
g_{11}^r(\omega) & =& \sum_{n=-\infty}^{\infty} \frac{w_{n}(1-n_0) + w_{-n} n_0}{\omega - (\varepsilon_d + n\omega_0) + i0^+}, \\
g_{22}^r(\omega) & =& \sum_{n=-\infty}^{\infty} \frac{w_{n}(1-n_0) + w_{-n} n_0}{\omega + (\varepsilon_d + n\omega_0) + i0^+},
\en
\end{subequations}
where $n_0$ represents the average occupation number on the QD, and $g_{12}^r(\omega)=g_{21}^r(\omega)=0$. The electron GF can be 
evaluated by the Dyson equation, $G_c=g+g (\Sigma_N^r +\Sigma_S^r) G_c$. Substituting the calculated $G_c$ into Eq.~(\ref{ar}), 
we can evaluate the Andreev reflection probability $T_A(\omega)$ under PTA. Therefore, we show in Fig.~\ref{figars} the 
comparison results of the DTA for $T_A(\omega)$ with those from SPA and PTA for the symmetric systems with $\varepsilon_d=0$, 
$0.1$, and $-0.2$, respectively, in equilibrium case and at zero temperature.

One can find from Fig.~\ref{figars} that the main feature of the DTA results is remarkably narrowing of the Andreev resonance in 
comparison with those of SPA, both of which show no phonon-assisted side peaks in the subgap region $|\omega|<\Delta$; On the 
contrary, the PTA results exhibit obvious subgap phonon side peaks at $\omega=\pm (\varepsilon_d+ n\omega_0)$ ($n\geq 0$) with 
nearly equal height.

For the electron-hole symmetric case $\varepsilon_d=\mu_S=0$, the Andreev reflection reaches resonance when the energy of 
incident electron is aligned with the Fermi energy of the SC electrode, i.e. $\omega=0$. Moreover, this central resonant peak 
exhibits remarkably rapid decrease for both DTA and PTA results, if the incident electron moves away from the resonant point.

In this QD system, the dot energy level can be tuned by applying gate voltage. When the dot level $\varepsilon_d$ is tuned away 
from the the symmetric point $\mu_S=0$, there are two Andreev resonant peaks locates at
$\omega=\pm |{\varepsilon}_d|$ for the SPA and DTA results, as the case of noninteracting system, but with a reduced amplitude. 
The situation may however be quite different for the PTA results. For the case of $\varepsilon_d=\pm 0.2$ (only the result of 
$\varepsilon_d=-0.2$ is plotted in Fig.~\ref{figars}(c)), the zero energy resonant peak is still remaining due to phonon assisted 
process, which is responsible for the appearance of the phonon side peak at the Andreev linear conductance by PTA as shown in 
Fig.~\ref{figac1} below.

Once again it should be emphasized that the PTA does not take into account vibronic dressing effect on electronic tunneling, 
while the more rigorous DTA indeed does, leading to complex dependence of the tunneling self-energies on bias voltage and 
temperature. That is the reason that the two methods predict great different Andreev reflection spectrum, which determines 
different exotic transport properties of the hybrid QD system at the subgap region, as we will see below.

\begin{figure}[htb]
\includegraphics[height=4cm,width=7.9cm]{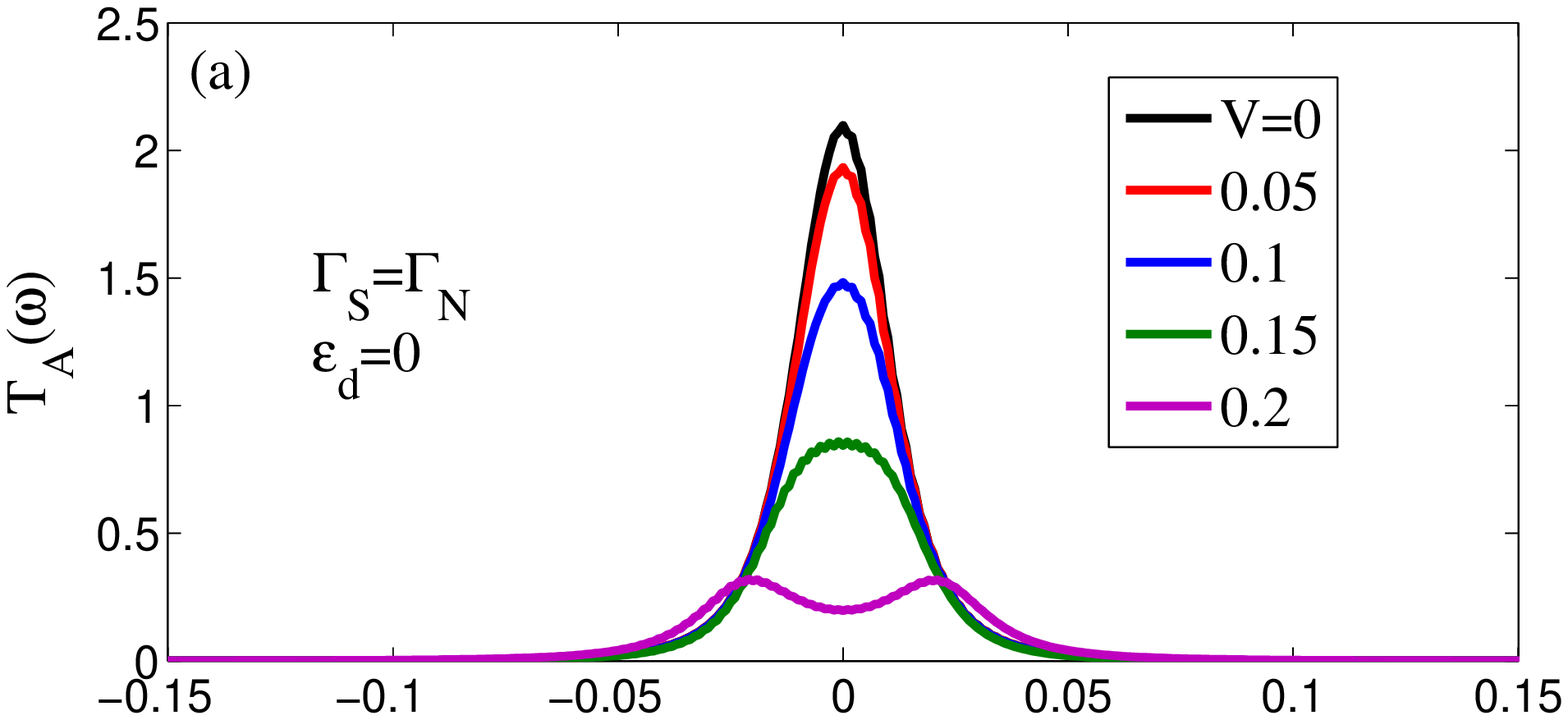}
\vspace{3mm}

\includegraphics[height=4cm,width=8cm]{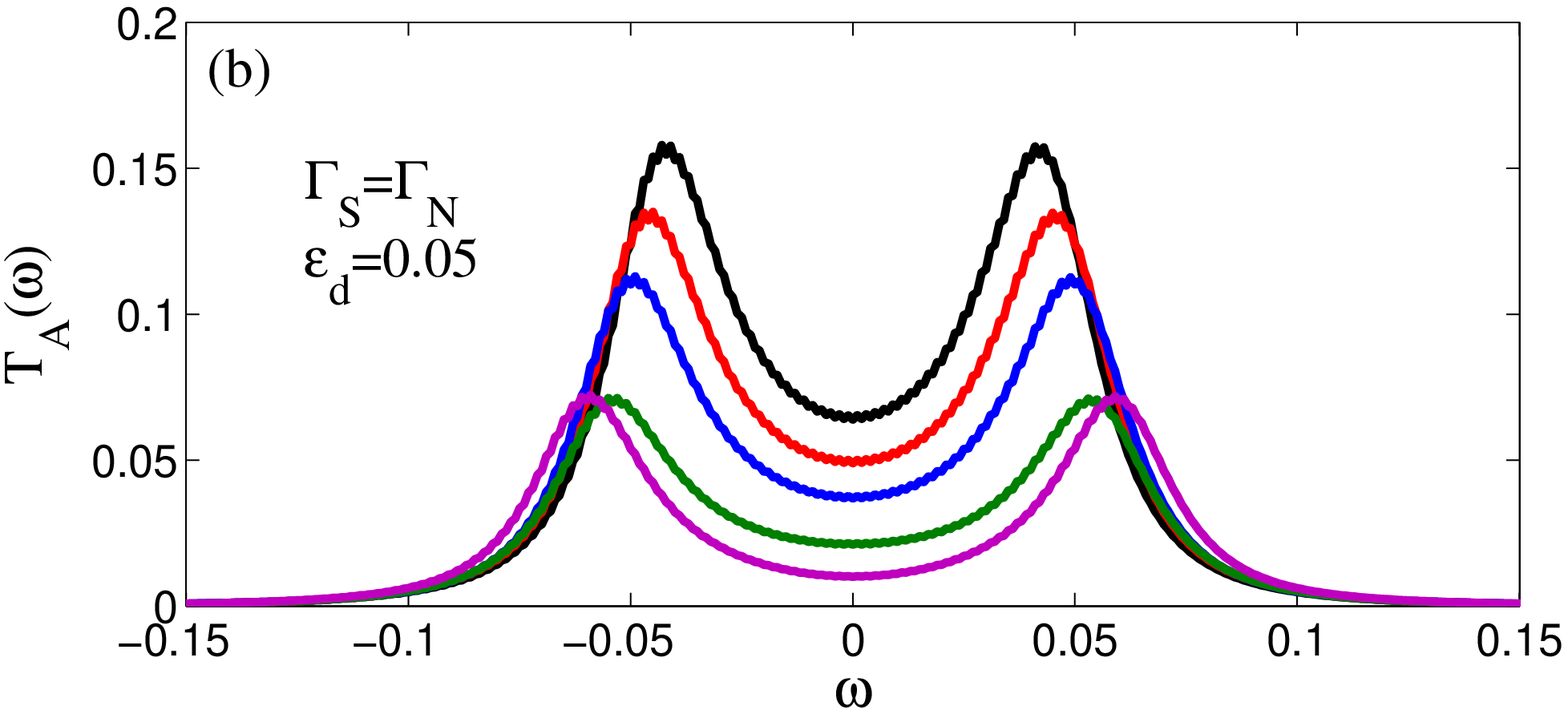}
\caption{(Colour online) The bias voltage dependence of the Andreev reflection spectrum in the DTA for the symmetric systems with 
$\varepsilon_d=0$ and $0.05$ at zero temperature.}
\label{figarsv}
\end{figure}

Moreover, we examine the bias voltage dependence of the Andreev reflection by DTA. As shown in Fig.~\ref{figarsv}, the resonant 
peak is slightly suppressed with increasing bias voltage, which can be ascribed to bias voltage dependent retarded self-energy. 
It is this suppression that results in a surprise decrease of the elastic tunneling current with increase of bias voltage.

\begin{figure}[htb]
\includegraphics[height=4cm,width=8cm]{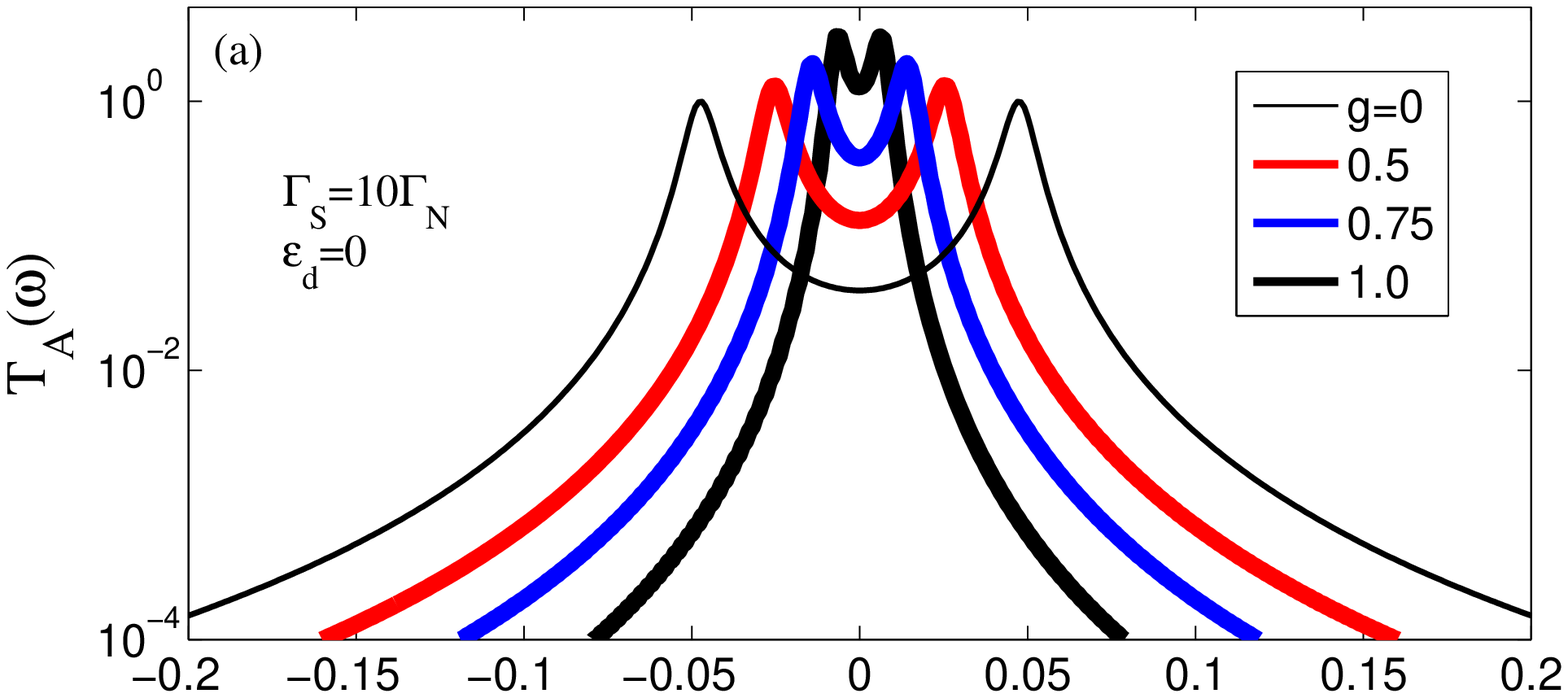}
\vspace{3mm}

\includegraphics[height=4cm,width=8cm]{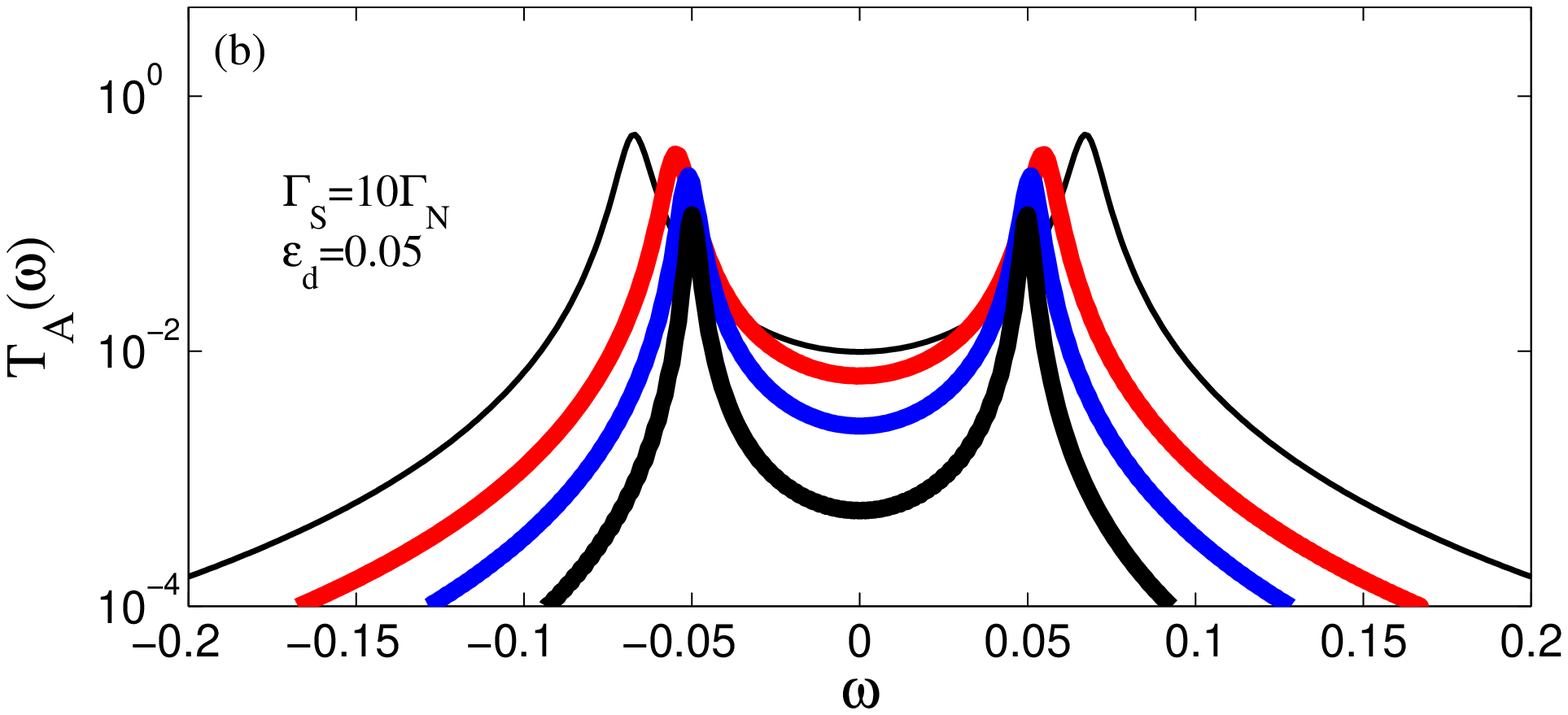}
\caption{(Colour online) The Andreev reflection spectrum in the DTA at zero bias voltage and zero temperature for the asymmetric 
N-QD-S system with $\Gamma_S=10\Gamma_N$, and different values of the EPI constant, $g=0$, $0.5$, $0.75$, and $1.0$. (a) is for 
the QD with the level $\varepsilon_d=0$, and (b) is for $\varepsilon_d=0.05$.}
\label{figarsay}
\end{figure}

Notice that the present theoretical method, DTA, can properly describe the dynamic properties of the EPI system not only in the 
strong coupling regime but also in the relatively weak coupling regime.\cite{Souto} At last of this section, we therefore analyze 
the Andreev reflection spectrum by DTA for the strongly asymmetric system with $\Gamma_S=10\Gamma_N$ and different EPI constants, 
$g=0.5$, $0.75$, and $1.0$. In Fig.~\ref{figarsay}, we show the DTA results for the Andreev reflection spectrum at these cases 
with $\varepsilon_d=0$ and $0.05$. The thin lines in this figure denote the corresponding results for the system without EPI. It 
is clear that owing to the strong SC proximity effect and weak coupling to the normal lead, two nearly resonant ingap states, 
i.e. the Andreev bound states, are distinctly emerging and cause subgap peaks in the Andreev reflection spectrum centered at 
energies $\pm \sqrt{\varepsilon_d^2+ \Gamma_S^2/4}$. The EPI effect is two folds. For the electron-hole symmetric system 
$\varepsilon_d=0$, inelastic scattering results in the resonant peaks shrinking progressively and becoming nearly single resonant 
peak with enhanced height at $\omega=0$ as EPI constant increases; while for the system $\varepsilon_d=0.05$, the resonant peaks 
are also gradually shrinking but eventually pinning at $\pm\varepsilon_d$ and a suppressed Andreev reflection spectrum at 
$\omega=0$.

\subsection{Linear Andreev conductance}

\begin{figure}[htb]
\includegraphics[height=4.5cm,width=8cm]{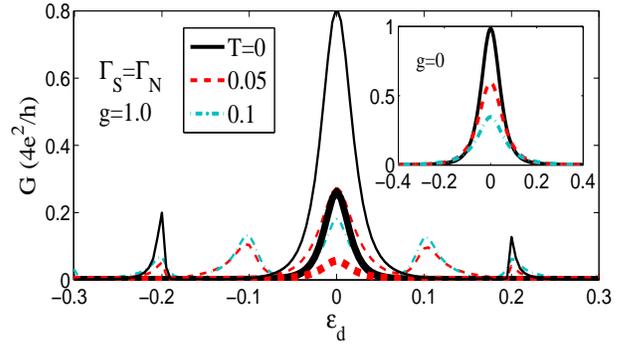}
\caption{(Colour online) The linear Andreev conductance $G$ for the symmetric system as a function of the resonant level 
$\varepsilon_d$ of the QD with $g=1$ at different temperatures $T=0$, $0.05$, and $0.1$. The thick lines correspond to the DTA 
results, while the thin lines to the PTA results. Inset: the corresponding results for the system in absence of EPI.}
\label{figac1}
\end{figure}

The DTA result for the linear Andreev conductance $G$ can be easily calculated from Eq.~(\ref{current}). At zero temperature, we 
have
\bn
G &=& \frac{dI_A}{dV}{\bigg |}_{V=0}=\frac{4e^2}{h} w_0^2|T_{A}(0)|^2 \cr
&=& \frac{4e^2}{h}\left\{ \frac{w_0 \Gamma_N \Re\Sigma_{c12}^r(0)}{\varepsilon_d^2 + [\Im\Sigma_{c11}^r(0)]^2+ 
[\Re\Sigma_{c12}^r(0)]^2} \right \}^2, \label{gae0}
\en
since $\Re\Sigma_{c11(22)}^r(0)=\Im \Sigma_{c12}^r(0)=0$, and $\Im\Sigma_{c11(22)}^r(0)=-iw_0 \Gamma_N /2$. While the PTA result 
for conductance can be simply obtained from Eq.~(\ref{gPTA}) as $G_{\rm PTA}=\frac{4e^2}{h} |T_{A}(0)|^2$ in the place of the 
Andreev reflection probability with the PTA one. We plot the two results as functions of dot level at different temperatures for 
the symmetric system in Fig.~\ref{figac1}.

For comparison, the Andreev conductance for noninteracting system is also plotted in Fig.~\ref{figac1}, which shows the following 
features: (1) the maximum value $G_0=4e^2/h$ for the resonant system $\varepsilon_d=0$ at zero temperature; (2) with increasing 
temperature, it is suppressed, but does not exhibit thermal broadening as the normal conductance does. For the EPI system, we 
observe a similar temperature behavior for the DTA results. Besides, the DTA predicts remarkable suppression of the Andreev 
conductance, $G=0.22G_0$, even for the resonant system $\varepsilon_d=0$ at zero temperature, which is ascribed to the large 
suppression of the nondiagonal element of the self-energy at the subgap region [see Fig.~\ref{figse2}(a)] and an extra $w_0^2$ 
factor in Eq.~(\ref{gae0}) due to the FC blockade.

More importantly, we find that the linear Andreev conductance $G$ exhibits no phonon side peaks as a function of the gate 
voltage, when $\varepsilon_d$ crosses the vibronic frequency. In specific, Eq.~(\ref{gae0}) shows that $G$ is proportional to 
$\varepsilon_d^{-4}$, indicating that the effect of the strong EPI is just to remarkably narrow the width of its resonance peak 
as the gate voltage is swept. This result is completely conflict with the PTA calculations, as indicated by the thin lines in 
Fig.~\ref{figac1}. For instance, the PTA calculations predicts two kinds of phonon assisted peaks: one is located at the phonon 
frequency,  $|\varepsilon_d|=\omega_0=0.2$, at zero temperature; the other one emerges at half of the phonon frequency, 
$|\varepsilon_d|=\omega_0/2=0.1$, with increasing temperatures. From mathematic point of view, the former one is stemming from 
the zero energy resonance in the Andreev reflection spectrum. While the later one results from the phonon assisted resonances of 
the Andreev reflection near $\omega=0$ at $\omega=\pm \omega_0/2=\pm 0.1$, as shown in Fig.~\ref{figars}(b), since these nearby 
peaks will make non-negligible contribution to integrals over energy in calculating linear-response conductance at sufficiently 
high temperature. It is noticed that albeit two peaks are also predicted at $\omega=\pm 0.1$ for the Andreev reflection spectrum 
by DTA, their heights are only two orders lower than those of PTA. They are therefore too weak to induce phonon side peak in 
linear transport.

Furthermore, we should notice that, as in the normal QD system, these appearance of phonon side peaks in linear response 
conductance is just an artifact of the used approximative method and is in fact unphysical. Since at zero bias voltage, the 
incident electron has at almost the same energy with the emitting electron, energy conservation does not allow the vibration to 
be excited during tunneling process. As in the normal QD system, the DTA corrects this artifact of the PTA by more precisely 
taking account of the vibrational effect on electronic tunneling self-energy.

\begin{figure}[htb]
\includegraphics[height=4.5cm,width=8cm]{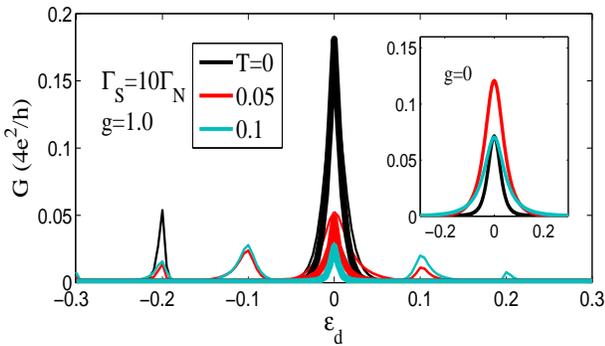}
\caption{(Colour online) The linear Andreev conductance $G$ for the asymmetric system $\Gamma_S=10\Gamma_N$. Other parameters are 
the same as those in Fig.~\ref{figac1}}
\label{figac2}
\end{figure}

Finally in this subsection, we investigate the Andreev conductance for the asymmetric system $\Gamma_S=10\Gamma_N$. In the same 
way, no phonon side peak is found for the DTA calculations as shown in Fig.~\ref{figac2}. At this case, the noninteracting system 
exhibits opposite temperature behavior to the symmetric system. The EPI effect abnormally enhances the central peak of the 
zero-temperature conductance due to the shrinking behavior at $\varepsilon_d=0$ in Fig.~\ref{figarsay}(a). As moving away from 
the electron-hole symmetric point, the EPI effect causes more rapidly narrowing of the peak of the conductance because of big 
suppression of the Andreev reflection spectrum at $\omega=0$ in Fig.~\ref{figarsay}(b).

\subsection{Andreev Current and differential conductance}

\begin{figure}[htb]
\includegraphics[height=5cm,width=8.5cm]{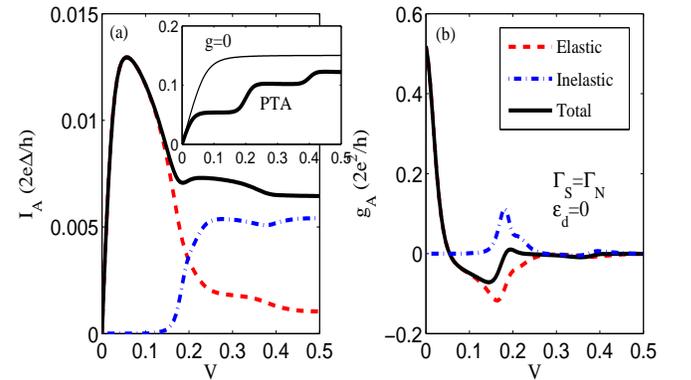}
\caption{(Colour online) (a) The calculated total current (solid line), elastic current (dashed line), inelastic current 
(dotted-dashed line); and (b) the corresponding differential conductances as functions of bias voltage for a symmetric system 
with $\varepsilon_d=0$ at zero temperature. The inset shows the PTA result (thick line) for the Andreev current and the current 
for noninteracting system (thin line).}
\label{figiv1}
\end{figure}

\begin{figure}[htb]
\includegraphics[height=5cm,width=8.5cm]{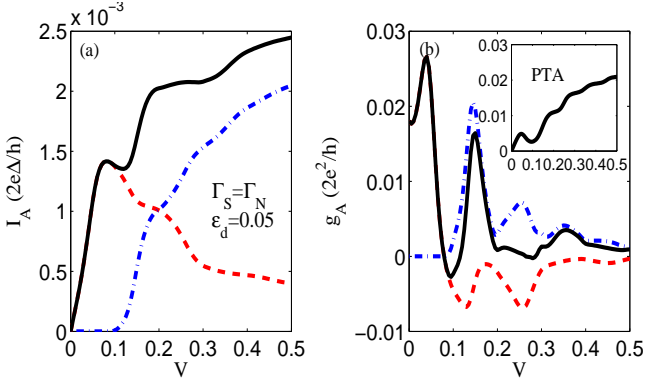}
\caption{(Colour online) The same figure as Fig.~\ref{figiv1} but for the system with $\varepsilon_d=0.05$.}
\label{figiv2}
\end{figure}

We now study the nonlinear transport at the subgap region. In Figs.~\ref{figiv1} and \ref{figiv2}, we plot the Andreev currents 
$I_A$ and corresponding differential conductances $g_A=dI_A/dV$ as functions of bias voltage $V>0$ for the symmetric systems with 
$\varepsilon_d=0$ and $0.05$, respectively, at zero temperature.
It would be very useful for understanding the calculated results by analyzing the respective contributions of the elastic and 
inelastic Andreev processes.
We therefore plot their corresponding elastic and inelastic contributions, and the total results as well.

At zero temperature, the elastic current formula Eq.~(\ref{currentel}) can be simplified as
\bq
I_{el}=\frac{2e}{h} w_0^2 \int_{- V}^{V} d\omega T_A(\omega). \label{currentel2}
\eq
The elastic current rises monotonously as usual at the beginning, and then suffers a decrease with increase of the bias voltage. 
Differentiating Eq.~(\ref{currentel2}) with respect to $V$, the elastic differential conductance can be written as two parts 
$g^{el}=dI_{el}/dV=g_{1}^{el}+g_{2}^{el}$, with
\bq
g_{1}^{el}= \frac{4e^2}{h} w_0^2 T_A(V) ,
\eq
and
\bq
g_{2}^{el} = \frac{2e^2}{h} w_0^2 \int_{-V}^{V} d\omega \frac{\partial T_A(\omega)}{\partial V} .
\eq
The first term, $g_{1}^{el}$, is proportional to the Andreev reflection $T_{A}(V)$ and results in a zero-bias maximum for the 
electron-hole symmetric system, and a nonzero-bias maximum, i.e. a resonant peak at $V=\varepsilon_d$, for the system with 
$\varepsilon_d=0.05$ due to peak splitting of the Andreev reflection spectrum as shown in Fig.~\ref{figarsv}(b); while the second 
term, $g_{2}^{el}$, is stemming from the bias voltage dependent Andreev reflection and makes negative contribution, which is 
responsible for decrease of the elastic current.

\begin{figure}[htb]
\includegraphics[height=3.5cm,width=8cm]{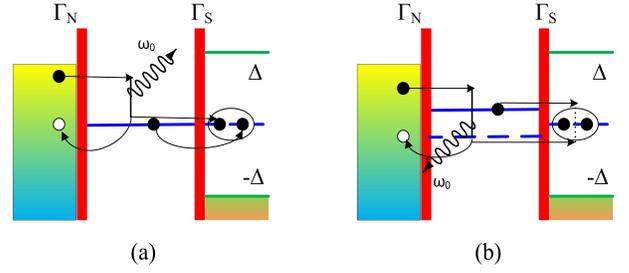}
\caption{(Colour online) Schematic diagram of the phonon emission assisted Andreev reflection processes involved in 
Figs.~\ref{figiv1} and \ref{figiv2}, respectively. Solid (open) circles denote the states of quasiparticles (quasiholes), and the 
arrows stand for directions of their tunneling. See text for more detail.}
\label{figsch}
\end{figure}

The inelastic current has a threshold at the onset for inelastic Andreev reflection processes, $V=\omega_0=0.2$, for the 
electron-hole symmetric system $\varepsilon_d=0$, for phonon emission which shows up as an abrupt increase of the inelastic 
current and a peak in the differential conductance as shown in Fig.~\ref{figiv1}. This observation can be interpreted, from 
mathematic point of view, by the dominant terms of the zero-temperature inelastic current
\bn
I_{in}&\simeq&  \frac{4e}{h} w_0 w_1 \left (\int_{\omega_0- V}^{V} + \int_{-V}^{V-\omega_0} \right ) d\omega T_A(\omega) \cr
&& + \frac{2e}{h} w_1^2 \int_{\omega_0- V}^{V-\omega_0} d\omega T_A(\omega).
\en
From physical perspective, such inelastic process can be illustrated in Fig.~\ref{figsch}(a): an electron with energy $\omega_0$ 
in the left normal lead can first emit a phonon to tunnel into the dot, and then enter the right SC lead to form a Cooper pair by 
picking up an electron originally in the dot, and finally reflect a hole back into the left lead.

More interestingly, as we consider the system moving away from the electron-hole symmetric case, i.e. $\varepsilon_d\neq 0$, we 
find a novel inelastic transport channel which is opening even at a bias voltage lower than the phonon energy. To be specific, 
Fig.~\ref{figiv2}(b) exhibits a relatively sharp peak at $V=0.15<\omega_0$ in the inelastic differential conductance for the 
system with $\varepsilon_d=0.05$. This peak can be ascribed to a new sort of phonon assisted Andreev reflection process as 
described in Fig.~\ref{figsch}(b): if the dot level is aligned at a positive energy, an electron with energy 
$\omega_0-\varepsilon_d$ in the left lead can first transverse the dot at the opposite energy level $-\varepsilon_d$ by emitting 
a phonon, and pick up an electron in the dot at the level $\varepsilon_d$ to form a Cooper pair. In this spirit, this sort of 
phonon assisted Andreev reflection can emerge at the condition $V=-|\varepsilon_d|+n \omega_0$ ($n> 0$). It is different from a 
N-QD-N system, where phonon assisted resonant tunneling will occur when the symmetrically applied bias voltage obeys the 
condition $V=2(|\varepsilon_d|+n\omega_0)$. Therefore, this newly predicted intriguing peak can be considered as a distinctive 
signature of the phonon assisted Andreev tunneling process. From Fig.~\ref{figiv2}(b), another peak in the inelastic current is 
found at $V=0.25$, which is stemming from the remaining effect of the peak splitting (its vibranic replica) in the Andreev 
reflection spectral function as shown in Fig.~\ref{figarsv}(b).

It is noticed that not only the inelastic Andreev current but also its elastic component exhibits phonon assisted features stated 
above since the opening of new inelastic processes will inevitably change the retarded self-energy. For instance, when the 
inelastic current channel is active, the term $g_2^{el}$ becomes predominant over the first term $g_1^{el}$, leading to further 
decrease of the elastic current and peaks (with negative values) of the elastic differential conductance. For the electron-hole 
symmetric system, $\varepsilon_d=0$, this behavior causes a negative differential conductance (NDC), which was observed in a 
recent experimental measurement.\cite{Gramich} However, the PTA result predicts completely different $I$-$V$ characteristics, no 
appearance of NDC, for the symmetric system.
For the electron-hole asymmetric system, $\varepsilon_d=0.05$, on the contrary, it is observed that the relatively strong 
inelastic peaks compensate the decrease of the elastic current and lead to nearly disappearance of NDC by DTA results. 
Nevertheless, the PTA calculation still exhibits an unambiguous NDC in the inset of Fig.~\ref{figiv2}(b).

\begin{figure}[htb]
\includegraphics[height=5cm,width=8.5cm]{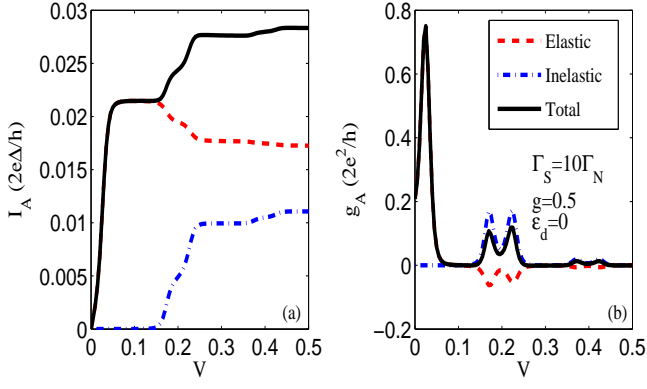}
\caption{(Colour online) The same figure as Fig.~\ref{figiv1} but for the asymmetric system with $\Gamma_S=10\Gamma_N$ and a weak 
EPI constant $g=0.5$.}
\label{figiv3}
\end{figure}

The $I$-$V$ characteristics of the asymmetric system $\Gamma_s=10\Gamma_N$ is shown in Fig.~\ref{figiv3} for the case of 
$\varepsilon_d=0$ and $g=0.5$. The subgap differential conductance $g_A$ shows a nonzero-bias maximum. This nonzero-bias anomaly 
has different origin from that of the symmetric system with $\varepsilon_d\neq 0$, which is stemming from the peak splitting of 
the Andreev reflection spectrum because of nonzero energy of the localized dot state. Here, the anomaly is due to distinct 
emergence of the ingap bound state, the Andreev state, in the extremely asymmetric tunnel-coupling. This bound state also induces 
peak splitting of the Andreev reflection spectrum as demonstrated in Fig.~\ref{figarsay}(a) as long as the EPI is not 
considerably strong. In addition, this peak splitting also affects the inelastic current, leading to a double-peak in the 
$g_A$-$V$ curve near the point $V=\omega_0=0.2$ where the inelastic channel is just opening.

\subsection{Zero-frequency shot noise}

In what follows, we analyze the zero-frequency shot noise at zero temperature, which can be calculated using a simplified 
expression according to the Eq.~(\ref{automf})
\bn
S_A &=& 2eI_A- \frac{4e^2}{h}\Gamma_N^2 \sum_{nmn'm'} w_n w_m w_{n'} w_{m'} \cr
&& \times \int_{\omega_1}^{\omega_2} d\omega T_A^2(\omega), \label{s0}
\en
with $\omega_1=\max(n\omega_0-V,n'\omega_0-V)$ and $\omega_2=\min(V-m\omega_0,V-m'\omega_0)$. We can also separate the shot noise 
as two contributions of elastic and inelastic parts, $S_A=S_{el}+S_{in}$, as
\begin{subequations}
\bn
S_{el}&=& 2eI_{el}- \frac{4e^2}{h}\Gamma_N^2 w_0^4 \int_{-V}^{V} d\omega T_A^2(\omega), \label{sT0el} \\
S_{in}&=& 2eI_{in}- \frac{4e^2}{h}\Gamma_N^2 w_0^2 (2w_0+w_1) w_1 \cr
&& \times\left (\int_{\omega_0- V}^{V} + \int_{-V}^{V-\omega_0} \right ) d\omega T_A^2(\omega) \cr
&& - \frac{16e^2}{h} \Gamma_N^2 w_0^2 w_1^2 \int_{\omega_0- V}^{V-\omega_0} d\omega T_A^2(\omega).  \label{sT0in}
\en
\end{subequations}

\begin{figure}[htb]
\includegraphics[height=5cm,width=8.5cm]{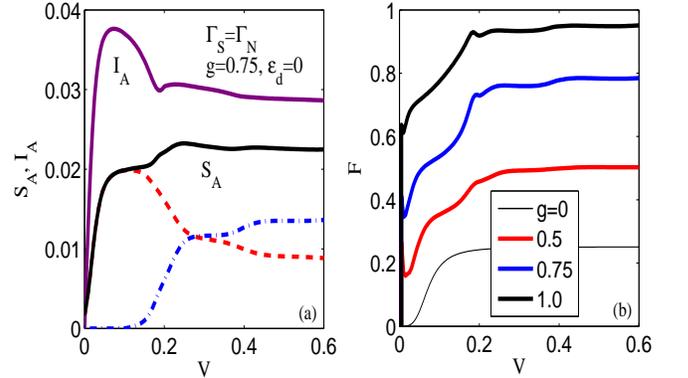}
\caption{(Colour online) (a) The zero-temperature shot noise (solid black line), and its elastic (dashed red line) and inelastic 
(dotted-dashed blue line) parts (normalized by $4e^2\Delta/h$) as functions of bias voltage for a symmetric N-QD-S with 
$\varepsilon_d=0$ and $g=0.75$. The total Andreev current is also plotted as a solid purple line; (b) The Fano factors for the 
system $\varepsilon_d=0$ with different EPI constants $g=1.0$ (black line), $0.75$ (blue line), and $0.5$ (red line). For 
comparison, the thin line denotes the Fano factor of the system without EPI.}
\label{fignv1}
\end{figure}

In Fig.~\ref{fignv1}(a), we plot the calculated shot noise, its two parts (normalized by $4e^2\Delta/h$), and the total current 
(normalized by $2e\Delta/h$) as functions of bias voltage $V>0$ for the symmetric system with $\varepsilon_d=0$ and $g=0.75$.
It is observed that for the QD $\varepsilon_d=0$, the elastic and inelastic noises inherit the same overall profiles as their 
corresponding currents with increasing bias voltage.
We also show the Fano factors, defined as
$F=S_A/e^*I_A$ (since the Andreev reflection corresponds to transferring twice the electron charge $e^*=2e$), in 
Fig.~\ref{fignv1}(b) for three values of the EPI parameters, $g=0.5$, $0.75$, and $1$. We find that the noise is greatly enhanced 
with increasing EPI strength. The Fano factor nearly approaches the value $1$ for the case of $g=1$ at high bias voltage region, 
while $F\approx 0.25$ for the noninteracting N-QD-S system. We can interpret the EPI-induced enhancement of noise as follows. The 
zero-temperature noise for the noninteracting system is simplified from Eq.~(\ref{sano}) as
\bq
S_{A}= 2eI_{A}- \frac{4e^2}{h}\Gamma_N^2 \int_{-V}^{V} d\omega T_A^2(\omega). \label{sT0no}
\eq
Because of the second term, the Fano factor is quite small. With increasing EPI strength, the contribution of the second term 
(and the third term) becomes more and more weaker due to the FC factors $w_0$ and $w_1$ in Eq.~(\ref{sT0el}) and (\ref{sT0in}) 
(FC blockade effect), for instance, $w_0=w_1=0.368$ at $g=1$. It is the FC blockade that induces enhancement of shot noise.

\begin{figure}[htb]
\includegraphics[height=5cm,width=8.5cm]{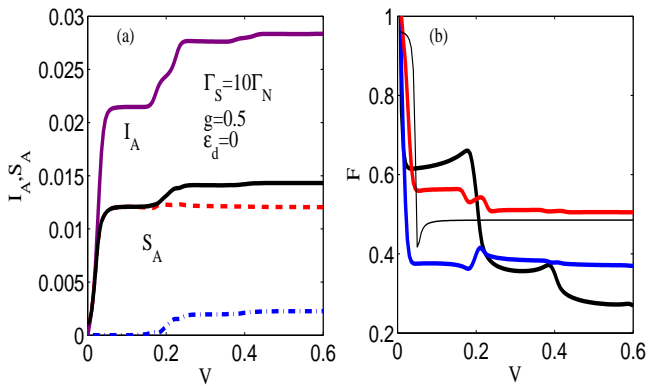}
\caption{(Colour online) This figure is the same as Fig.~\ref{fignv1} except for the asymmetric system with 
$\Gamma_S=10\Gamma_N$, and $g=0.5$ in (a).}
\label{fignv2}
\end{figure}

For the asymmetric tunnel-coupling case considered in the present paper, $\Gamma_N=10\Gamma_S$, the situation is quite different 
as illustrated in Fig.~\ref{fignv2}.
First of all, the noninteracting system has a greater Fano factor, $F\approx 0.5$, than that of the symmetric system since the 
second term in Eq.~(\ref{sT0no}) has a smaller contribution due to the peak splitting in the Andreev reflection spectrum with 
suppressed magnitude as demonstrated in Fig.~\ref{figarsay}(a). However, the splitting peaks will become shrinking and enhanced 
with increasing EPI strength, even become probably a single resonant peak at the case of $g=1$. As a result, the contribution of 
the second and third terms in Eqs.~(\ref{sT0el}) and (\ref{sT0in}) becomes gradually more and more important with increasing EPI 
strength, in spite of the FC factor. Consequently, strong EPI could lead to suppression of shot noise at large bias voltage 
region, where the inelastic channel is opening. For instance, $F\approx 0.27$ at $V=0.6$ for the system with $g=1$ in 
Fig.~\ref{fignv2}(b). While if the inelastic channel is not opening at small bias voltage region ($V\leq 0.2$ for the present 
case), only the second term in Eq.~(\ref{sT0el}) is necessary to be considered. This fact will therefore result in an enhancement 
of noise for the case of $g=1$ due to FC blockade effect $w_0=0.368$, but still a suppression for the case of $g=0.75$ due to a 
bigger $w_0=0.57$.

\section{Conclusion}

In conclusion, we have present a theory of the FCS of vibrational assisted electronic tunneling through a hybrid N-QD-S system in 
the subgap region on the basis of the Lang-Firsov canonical transformation for the local EPI and NGF method. In order to examine 
the interplay between the vibrational modified tunneling and the SC proximity effect, we have generalized the DTA decoupling 
scheme, which has been recently developed to study vibrational assisted stationary and transient tunneling in the N-QD-N system, 
to the hybrid system. Since this method takes more rigorous account of the phononic propagator in evaluation of the electronic 
GF, it is believed to provide correct description for phonon assisted electronic tunneling in the polaronic regime and even the 
crossover regime, $g^2\omega_0\leq \Gamma_N$, by comparison with other methods, for example, the SPA and the PTA. Besides, the 
DTA can provide an explicit analytical formula for the cumulant generating function for the vibrational assisted Andreev 
tunneling, and give analytical expressions for the current, zero-frequency shot noise, and etc.

In the first step we have discussed in detail phonon effects on the Andreev reflection spectrum for the symmetric and asymmetric 
hybrid systems and their bias voltage dependence, since it is the underlying sole physical quantity to characterize subgap 
transport properties of a hybrid N-QD-S system. Different from the PTA results for the subgap Andreev reflection spectrum, the 
DTA results have predicted no appearance of phonon assisted side peaks, instead a remarkably narrowed single peak and/or 
splitting peak in the subgap region.

In the second step we have investigated inelastic effects on the subgap $I$-$V$ characteristics. The PTA calculations predict 
that the linear Andreev conductance $G$ will show phonon assisted side peaks as the gate voltage is swept. Noticeably, one of the 
most important results in the present studies is to correct the unphysical result and predict a single narrowed peak in the 
$G$-$\varepsilon_d$ curves. Another intriguing result is about inelastic effects on the nonlinear transport. Two kinds of phonon 
assisted Andreev reflection processes are identified for the electron-hole symmetric system $\varepsilon_d=0$ and the asymmetric 
system $\varepsilon_d\neq 0$, respectively. In particular, a novel upward step in current is found for the latter case, showing a 
resonant peak in its differential conductance $g$-$V$ curve, when the inelastic Andreev reflection channel is opening at $V=-\mid 
\varepsilon_d \mid + \omega_0$. It is noticed that this phonon assisted resonant tunneling condition, $V<\omega_0$, is quite 
different from that of normal electronic tunneling at N-QD-N system. It is therefore can be regarded as a representative 
signature of the vibronic assisted Andreev reflection. Moreover, for the former case a pronounced decrease of the elastic part of 
the current is observed with increasing bias voltage, leading to a NDC at small bias voltage region. We have pointed out that 
this NDC is stemming from the bias-voltage induced suppression of the Andreev reflection spectrum.

In the final step we have examined the zero-frequency shot noise at zero temperature. By analyzing the Fano factor, we have 
displayed that vibronic effect could induce either enhancement of noise in the electron-hole symmetric case or suppression of 
noise in the electron-hole asymmetric case at large bias voltage limit.

Some of these predictions, for instant, no appearance of phonon assisted side peak in the linear conductance and a NDC behavior 
in the nonlinear current at the subgap region, are in good qualitative agreement with recent experimental measurement on 
inelastic Andreev tunneling on a Carbon nanotube QD. We hope that our other findings could be tested in the future experiments.

\begin{acknowledgments}

This work was supported by Projects of the National Basic Research Program of China (973 Program) under Grant No. 2011CB925603, 
and the National Science Foundation of China, Specialized Research Fund for the Doctoral Program of Higher Education (SRFDP) of 
China.

\end{acknowledgments}


\begin{thebibliography}{99}

\bibitem{hPark}{H. Park, J. Park, A. Lim, E. Anderson, A. Allvisatos, and P. McEuen, Nature (London) {\bf 407}, 57 (2000).}

\bibitem{jPark}{J. Park, A.N. Pasupathy, J.I. Goldsmith, C. Chang, Y. Yaish, J.R. Petta, M. Rinkoski, J.P. Sethna, H. Abruna, 
P.L. McEuen, and D.C. Ralph, Nature {\bf 417}, 722 (2002); N.B. Zhitenev, H. Meng, and Z. Bao, Phys. Rev. Lett. {\bf 88}, 226801 
(2002).}

\bibitem{Weig}{E.M. Weig, R.H. Blick, T. Brandes, J. Kirschbaum, W. Wegscheider, M. Bichler, and J.P. Kotthaus, Phys. Rev. Lett. 
{\bf 92}, 046804 (2004); L.H. Yu, Z.K. Keane, J.W. Ciszek, L. Cheng, M.P. Stewart, J.M. Tour, and D. Natelson, Phys. Rev. Lett. 
{\bf 93}, 266802 (2004); L.H. Yu and D. Natelson, Nano Lett. {\bf 4}, 79 (2004); A.N. Pasupathy, J. Park, C. Chang, A.V. 
Soldatov, S. Lebedkin, R.C. Bialczak, J.E. Grose, L.A.K. Donev, J.P. Sethna, D.C. Ralph, and P.L. McEuen, Nano Lett. {\bf 5}, 203 
(2005).}

\bibitem{LeRoy}{B.J. LeRoy, S.G. Lemay, J. Kong, and C. Dekker, Nature {\bf 432}, 371 (2004); B.J. LeRoy, J. Kong, V.K. 
Pahilwani, C. Dekker, and S.G. Lemay, Phys. Rev. B {\bf 72}, 075413 (2005); S. Sapmaz, P. Jarillo-Herrero, Ya.M. Blanter, C. 
Dekker, and H.S.J. van der Zant, Phys. Rev. Lett. {\bf 96}, 026801 (2006).}

\bibitem{Mitra}{A. Mitra, I. Aleiner, and A.J. Millis, Phys. Rev. B {\bf 69}, 245302 (2004); Phys. Rev. Lett {\bf 94}, 076404 
(2005).}

\bibitem{Koch}{J. Koch and F. von Oppen, Phys. Rev. Lett. {\bf 94}, 206804 (2005); J. Koch and F. von Oppen, Phys. Rev. B {\bf 
72}, 113308 (2005); J. Koch, M.E. Raikh, and F. von Oppen, Phys. Rev. Lett. {\bf 95}, 056801 (2005).}

\bibitem{Zazunov}{A. Zazunov, D. Feinberg, and T. Martin, Phys. Rev. B {\bf 73}, 115405 (2006).}

\bibitem{Shen}{X.Y. Shen, B. Dong, X.L. Lei, and N.J.M. Horing, Phys. Rev. B {\bf 76}, 115308 (2007); B. Dong, X.L. Lei, and 
N.J.M. Horing, IEEE Sensor Journal {\bf 8}, 885 (2008); B. Dong, H.Y. Fan, X.L. Lei, and N.J.M. Horing, J. Appl. Phys. {\bf 105}, 
113702 (2009).}

\bibitem{Frederiksen}{T. Frederiksen, M. Brandbyge, N. Lorente, and A.P. Jauho, Phys. Rev. Lett. {\bf 93}, 256601 (2004); M. 
Paulsson, T. Frederiksen, and M. Brandbyge, Phys. Rev. B {\bf 72}, 201101 (2005); T. Frederiksen, M. Paulsson, M. Brandbyge, and 
A.P. Jauho, Phys. Rev. B {\bf 75}, 205413 (2007).}

\bibitem{Vega}{L. de la Vega, A. Mart\'in-Rodero, N. Agra\"it, and A. Levy Yeyati, Phys. Rev. B {\bf 73}, 075428 (2006).}

\bibitem{Viljas}{J.K. Viljas, J.C. Cuevas, F. Pauly, and M. H\"afner, Phys. Rev. B {\bf 72}, 245415 (2005); R. Egger and A.O. 
Gogolin, Phys. Rev. B {\bf 77}, 113405 (2008)}

\bibitem{Entin}{O. Entin-Wohlman, Y. Imry, and A. Aharony, Phys. Rev. B {\bf 80}, 035417 (2009).}

\bibitem{Haule}{K. Haule and J. Bon\v ca, Phys. Rev. B {\bf 59}, 13087 (1999); E.G. Emberly and G. Kirczenow, Phys. Rev. B {\bf 
61}, 5740 (2000); B. Dong, H.L. Cui, and X.L. Lei, Phys. Rev. B {\bf 69}, 205315 (2004); B. Dong, H.L. Cui, X.L. Lei, and N.J.M. 
Horing, Phys. Rev. B {\bf 71}, 045331 (2005).}

\bibitem{Flensberg}{K. Flensberg, Phys. Rev. B {\bf 68}, 205323 (2003).}

\bibitem{Rodero}{A. Martin-Rodero, A. Levy Yeyati, F. Flores, and R.C. Monreal, Phys. Rev. B {\bf 78}, 235112 (2008); R.C. 
Monreal and A. Martin-Rodero, Phys. Rev. B {\bf 79}, 115140 (2009); R.C. Monreal, F. Flores, and A. Martin-Rodero, Phys. Rev. B 
{\bf 82}, 235412 (2010).}

\bibitem{Galperin}{M. Galperin, A. Nitzan, and M.A. Ratner, Phys. Rev. B {\bf 73}, 045314 (2006); M. Galperin, A. Nitzan, and 
M.A. Ratner, Phys. Rev. B {\bf 74}, 075326 (2006); R. H\"artle, C. Benesch, and M. Thoss, Phys. Rev. B {\bf 77}, 205314 (2008); 
R. H\"artle, M. Butzin, O. Rubio-Pons, and M. Thoss, Phys. Rev. Lett. {\bf 107}, 046802 (2011).}

\bibitem{Zazunov2}{A. Zazunov and T. Martin, Phys. Rev. B {\bf 76}, 033417 (2007).}

\bibitem{Schmidt}{T.L. Schmidt, and A. Komnik, Phys. Rev. B {\bf 80}, 041307 (2009); R. Avriller and A. Levy Yeyati, Phys. Rev. B 
{\bf 80}, 041309 (2009); F. Haupt, T. Novotn\'y, and W. Belzig, Phys. Rev. Lett. {\bf 103}, 136601 (2009); F. Haupt, T. 
Novotn\'y, and W. Belzig, Phys. Rev. B {\bf 82}, 165441 (2010).}

\bibitem{Avriller}{R. Avriller and T. Frederiksen, Phys. Rev. B {\bf 86}, 155411 (2012).}

\bibitem{Maier}{S. Maier, T.L. Schmidt, and A. Komnik, Phys. Rev. B {\bf 83}, 085401 (2011).}

\bibitem{Dong}{B. Dong, G.H. Ding, and X.L. Lei, Phys. Rev. B {\bf 88}, 075414 (2013).}

\bibitem{Souto}{R. Seoane Souto, A.L. Yeyati, A. Martin-Rodero, and R.C. Monreal, Phys. Rev. B {\bf 89}, 085412 (2014).}

\bibitem{Hofstetter2}{L. Hofstetter, A. Geresdi, M. Aagesen, J. Nyg\aa rd, C. Sch\"onenberger, and S. Csonka, Phys. Rev. Lett. 
{\bf 104}, 246804(2010).}

\bibitem{Deacon}{R. S. Deacon, Y. Tanaka, A. Oiwa, R. Sakano, K. Yoshida, K. Shibata, K. Hirakawa, and S. Tarucha, Phys. Rev. 
Lett. {\bf 104}, 076805 (2010); Phys. Rev. B {\bf 81}, 121308 (2010).}

\bibitem{Franke}{K.J. Franke, G. Schulze, and J.J. Pascual, Science {\bf 332}, 940 (2011).}

\bibitem{Pillet}{J.-D. Pillet, C.H.L. Quay, P. Morfin, C. Bena, A. Levy Yeyati, and P. Joyez, Nat. Phys. {\bf 6}, 965 (2010);
T. Dirks, T.L. Hughes, S. Lal, B. Uchoa, Y.-F. Chen, C. Chialvo, P.M. Goldbart, and N. Mason, Nat. Phys. {\bf 7}, 386 (2011); 
E.J.H. Lee, X. Jiang, M. Houzet, R. Aguado, C.M. Lieber, and S. De Franceschi, Nat. Nanotechnol. {\bf 9}, 79 (2013).}

\bibitem{Hofstetter}{L. Hofstetter, S. Csonka, J. Nyg\aa rd, and C. Sch\"onenberger, Nature (London) {\bf 461}, 960 (2009); L.G. 
Herrmann, F. Portier, P. Roche, A.L. Yeyati, T. Kontos, and C. Strunk, Phys. Rev. Lett. {\bf 104}, 026801 (2010); L. Hofstetter, 
S. Csonka, A. Baumgartner, G. F\"ul\"op, S. d'Hollosy, J. Nyg\aa rd, and C. Sch\"onenberger, Phys. Rev. Lett. {\bf 107}, 136801 
(2011); J. Schindele, A. Baumgartner, and C. Sch\"onenberger, Phys. Rev. Lett. {\bf 109}, 157002 (2012).}

\bibitem{Gramich}{J. Gramich, A. Baumgartner, and C. Sch\"onenberger, Phys. Rev. Lett. {\bf 115}, 216801 (2015).}

\bibitem{Zhang2}{P. Zhang and Y.-X. Li, J. Phys.: Condens. Matter {\bf 21}, 095602 (2009).}

\bibitem{Golez}{D. Gole\v z, J. Bon\v ca, and R. \v Zitko, Phys. Rev. B {\bf 86}, 085142 (2012).}

\bibitem{Zhang}{S.N. Zhang, W. Pei, T.F. Fang, and Q.F. Sun, Phys. Rev. B {\bf 86}, 104513 (2012).}

\bibitem{Bocian}{K. Bocian and W. Rudzi\'nski, Eur. Phys. J. B {\bf 88}, 50 (2015).}

\bibitem{Baranski}{J. Bara\'nski and T. Doma\'nski, J. Phys.: Condens. Matter {\bf 27}, 305302 (2015).}

\bibitem{Stadler}{P. Stadler, W. Belzig, and G. Rastelli, Phys. Rev. Lett. {\bf 117}, 197202 (2016).}

\bibitem{Gogolin}{A. Komnik and A.O. Gogolin, Phys. Rev. Lett. {\bf 94}, 216601 (2005); A.O. Gogolin and A. Komnik, Phys. Rev. B 
{\bf 73}, 195301 (2006).}

\bibitem{Mahan}{G.D. Mahan, {\em Many-Particle Physics.} (Third edition, Kluwer Academic/Plenum Publisher, New York, 2000).}

\end{thebibliography}
\end{document}